\UseRawInputEncoding 
\documentclass[journal,onecolumn]{IEEEtran}
\usepackage{graphicx}
\usepackage{amsmath,amssymb}
\usepackage[colorlinks=true, pdfstartview=FitV, linkcolor=blue, citecolor=blue, urlcolor=blue]{hyperref}
\usepackage{float}
\restylefloat{table}
\usepackage[english, footnote, nomargin, draft, marginclue]{fixme} 
\usepackage{physics}
\usepackage{mathtools}
\allowdisplaybreaks

\usepackage{ulem} 
\usepackage{xcolor} 

\begin{document}

\bstctlcite{BSTcontrol}

\title{To Be a Truster or Not to Be: Evolutionary Dynamics of a Symmetric N-Player Trust Game in Well-Mixed and Networked Populations
}

\author{Ik~Soo~Lim and Naoki~Masuda
\thanks{Ik~Soo~Lim  is with 
School of Computing and Mathematical Sciences, University of Greenwich, London SE10 9LS, U.K. 
(e-mail: i.lim@gre.ac.uk, iksoolim@gmail.com).
\\
Naoki~Masuda is with 
Department of Mathematics and Institute for Artificial Intelligence and Data Science, 
State University of New York at Buffalo, USA.
}
\thanks{This work has been submitted to the IEEE for possible publication. Copyright may be transferred without notice, after which this version may no longer be accessible.}
}

\markboth{
}%
{Shell \MakeLowercase{\textit{et al.}}: Bare Demo of IEEEtran.cls for IEEE Journals}

\maketitle



\begin{abstract}

Trust and reciprocation of it form the foundation of economic, social and other interactions.
While the Trust Game is widely used to study these concepts for interactions between two players, often alternating different roles (i.e., investor and trustee), its extensions to multi-player scenarios have been restricted to instances where players assume only one role.  
We propose a symmetric N-player Trust Game, in which players alternate between two roles, and the payoff of the player is defined as the average across their two roles and drives the evolutionary game dynamics. We find that prosocial strategies are harder to evolve with the present symmetric N-player Trust Game than with the Public Goods Game, which is well studied. 
In particular, trust fails to evolve regardless of payoff function nonlinearity in well-mixed populations in the case of the symmetric N-player trust game. 
In structured populations, nonlinear payoffs can have strong impacts on the evolution of trust.
The same nonlinearity can yield substantially different outcomes, depending on the nature of the underlying network. Our results highlight the importance of considering both payoff structures and network topologies in understanding the emergence and maintenance of prosocial behaviours.

\end{abstract}


%
\IEEEpeerreviewmaketitle

\section{Introduction} 
 
\subsection{Evolution of Prosocial Behaviours}  

Researchers across disciplines have explored how prosocial behaviours evolve among self-interested individuals. A key focus is the evolution of cooperation in social dilemmas, such as the Prisoner's Dilemma (PD) game and its $N$-player variant, the Public Goods Game (PGG) 
\cite{
Zeng:2025fv, Feng:2024lq, Mathieu:2023fk, Guo:2023qf, 
nowak2006five, hauert2006synergy, Van-Segbroeck:2009uq, Sasaki:2012rz}.
In particular, evolutionary game theory enables us to examine how successful strategies proliferate through evolution, i.e., fitness-dependent reproduction and imitation \cite{SMITH:1973aa, Taylor:1978aa}. 
Evolutionary game theory has also been employed to investigate information dynamics in evolving networks, as well as cooperative packet forwarding in mobile ad hoc networks \cite{Feng:2024vn, tang2015when}.

\begin{figure}[t!]  
\begin{center}  
\includegraphics[width=0.3\textwidth]{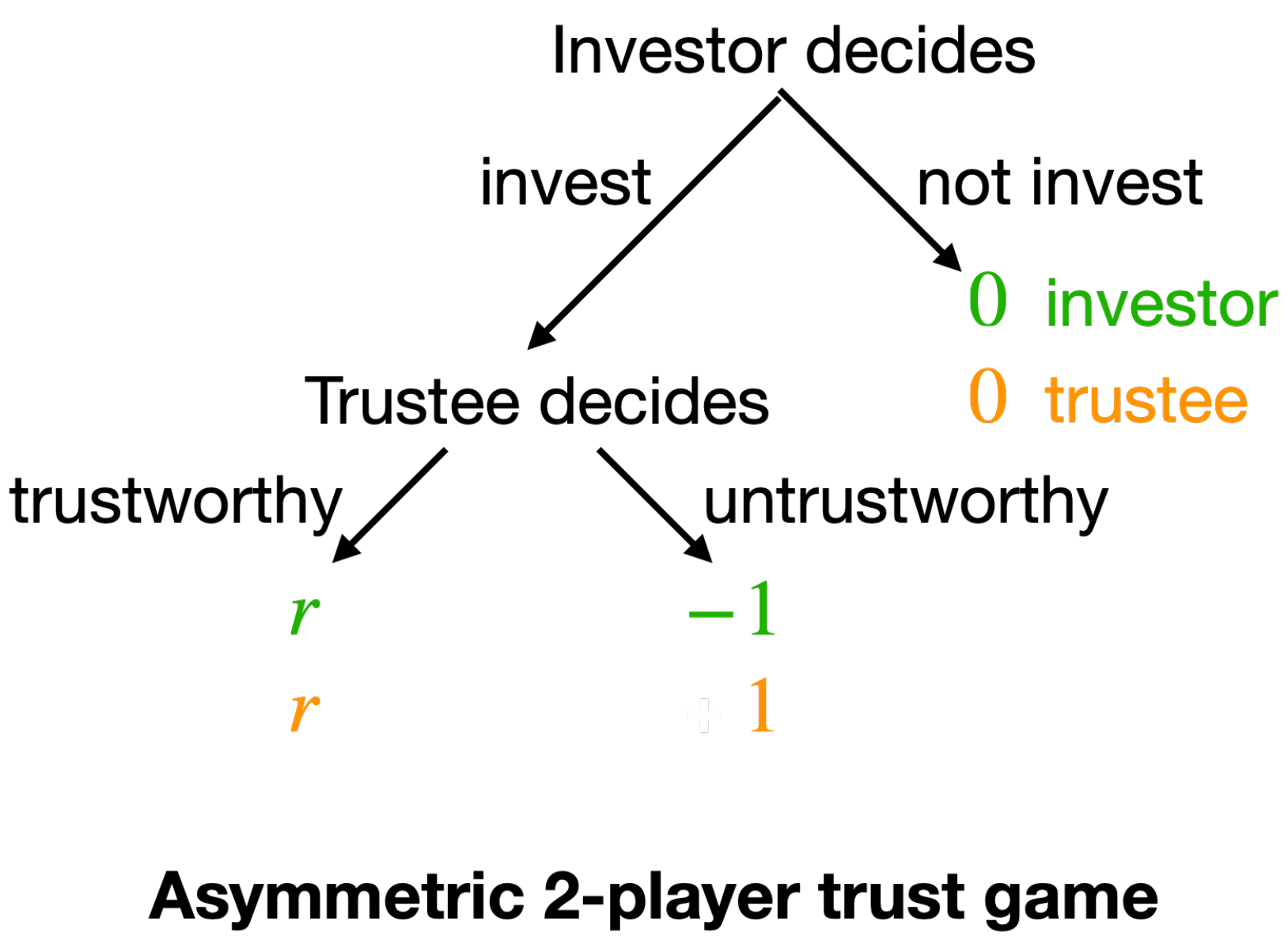}             
\end{center}  
\caption{
Game tree of the asymmetric 2-player binary TG,
in which the role of each player is fixed. 
The payoffs of an investor are shown in green. 
Those of a trustee are shown in orange.
We require $0<r<1$, where $r$ represents the relative productivity of the prosocial strategies. 
Adapted from Ref.~\cite{Masuda:2012aa}.  
}    
\label{fig_trust_game}  
\end{figure}

Many real-world scenarios involve sequential interactions between two players, as seen in buyer-seller exchanges. Unlike the PD and PGG, which model simultaneous interactions, sequential interactions introduce a trust issue where one player's decision may leave them vulnerable to exploitation by the other \cite{Bravo:2008aa}. 
The concepts of trust and trustworthiness have also gained attention in engineering research \cite{Li:2023kq, Calegari:2024aa, Ting:2021rc, Braga:2018qf, Cho:2015aa}. 
Many problems in these fields are framed as buyer-seller interactions \cite{Jung:2019yq}.
The Trust Game (TG) is a current standard for formalising non-simultaneous interactions under social dilemmas and is widely used for studying trust and trustworthiness \cite{Bravo:2008aa, Johnson:2011aa, Camerer:1988fk, Masuda:2012aa, McNamara:2009aa, Kumar:2020aa}. The TG involves a one-shot sequential interaction between an investor (truster)  and a trustee. 
The binary TG is a simple variant where each role has two strategies: investors can choose to invest or not, and trustees can choose to be trustworthy or untrustworthy (Fig.\,\ref{fig_trust_game}) \cite{Masuda:2012aa, Bohnet:2004xe, Guth:2000qq}.
Evolutionary game theory predicts that anti-social strategies evolve in the 2-player TG, resulting in reduced payoffs for both players compared to prosocial strategies.
Consequently, an additional mechanism is needed for prosocial strategies to evolve in the TG \cite{Manapat:2013aa, Tarnita:2015aa, Lim:2022vt}.
In practice, interactions often involve multiple participants, so recent attention has been paid to $N$-player TG (NTG) \cite{Abbass:2016aa, Chica:2018wt, Hu:2021tl, Fang:2021ut, Lim:2024ty,Zhou:2024zr}, which attempts to generalise the 2-player TG to multiplayer settings, with $N\geq 3$.

\subsection{Existing NTG Models and Rationale for Symmetry}

Two alternative approaches exist for NTG, each with distinct numbers of strategies and imitation processes.

\subsubsection{Original NTG}
        
The original NTG model employs three strategies: one for investors and two for trustees \cite{Abbass:2016aa}. In this model, investors can only invest, and trustees can choose to be trustworthy or untrustworthy. The evolutionary game dynamics are driven by role-independent imitation of strategies, wherein an investor can imitate a trustee's strategy and thus become a trustee, and vice versa. Subsequent NTG studies often adopt these assumptions \cite{Chica:2018wt,Hu:2021tl, Fang:2021ut, Sun:2022uq, Liu:2022fk, Xia:2022yq, Li:2023kx, Feng:2023fr,Liu:2024db}. Without additional mechanisms, the evolutionary outcome in a well-mixed population is the extinction of investors, with only trustees surviving. The lack of investment predicted by classical game theory aligns with the absence of surviving investors derived by this version of evolutionary game theory.

\subsubsection{Asymmetric NTG}

An alternative approach is an asymmetric NTG model that employs four strategies, two per role \cite{Lim:2024ty}. In this model, investors can choose to invest or not. The evolutionary game dynamics are driven by role-dependent strategy imitation between investors or trustees, but not between an investor and a trustee. In this asymmetric NTG, players maintain fixed roles throughout the evolutionary process.
Without additional mechanisms, the evolutionary outcome in a well-mixed population is the evolution of non-investing investors.
The lack of investment predicted by classical game theory is consistent with the evolution of investors who do not invest in this version of evolutionary game theory.

Existing NTG models, including the asymmetric NTG detailed previously, typically assume fixed roles with evolutionary dynamics driven by single-role payoffs. However, real-world interactions often involve fluid roles. Even in the standard 2-player TG context, which is inherently asymmetric, symmetrisation (where players alternate between investor and trustee roles) is well-established \cite{sigmund2016calculus, Lim:2022vt}. This approach reflects the common real-world occurrence of role alternation; for instance, individuals often act as both buyer (investor) and seller (trustee), suggesting that incorporating this dynamic is a worthwhile consideration for modelling trust dynamics.

The relevance of role alternation arguably extends to complex, multi-player settings requiring collective action between groups. Consider, for instance, joint ventures or supply chain clusters where enterprises form a buying consortium (acting as multiple buyers/investors) to jointly buy materials or capacity, needing collective action for scale or access to specialised suppliers. This group of buyers decides whether to collectively engage with a selling alliance (acting as multiple sellers/trustees). Engaging involves placing orders (an act of investment) and implicitly trusting the seller group to act trustworthily, for example, by meeting quality and delivery standards, rather than acting untrustworthily. Crucially, participants often switch roles: the same enterprise in the buying consortium might simultaneously participate in a separate selling alliance (acting as one of multiple sellers/trustees) supplying components to another buying consortium.
Such $N$-player scenarios, involving group interactions where entities operate as part of both investor groups and trustee groups, further highlight the potential limitations of fixed-role models. The frequent use of role alternation in behavioural TG experiments also points towards its perceived importance \cite{Burks:2003vn,Altmann:2008uq,Espin:2016fk}.

Given these motivations for considering models with role alternation, a key question arises: how should evolutionary fitness be determined when individuals occupy multiple roles over time? If players frequently switch between being investors and trustees, their overall success likely depends on their performance across both activities. Indeed, existing models of symmetric 2-player TGs often address this by using the average payoff across roles to determine fitness \cite{sigmund2016calculus, Lim:2022vt}. Evaluating fitness based solely on performance in a single, temporarily held role may therefore be insufficient in dynamic settings. Extending this logic to the multi-player context, we propose a symmetric NTG (SNTG) where evolutionary fitness is explicitly determined by the average payoff accumulated across both the investor and trustee roles. This mechanism contrasts fundamentally with the single-role payoff calculations underpinning fixed-role models or the imitation dynamics in the original NTG.

This paper formally defines this SNTG framework, motivated by the prevalence of role-switching and based on an average payoff structure analogous to that used in symmetric 2-player models, and investigates its evolutionary dynamics. We particularly explore how trust evolves within various population structures and interaction conditions under this symmetric approach, where prosocial outcomes depend on agents successfully navigating both the investor role (e.g., choosing to trust/invest) and the trustee role (e.g., choosing to be trustworthy).

\subsection{Contributions}

\begin{figure}[t!]  
\begin{center}  
\includegraphics[width=0.49\textwidth]{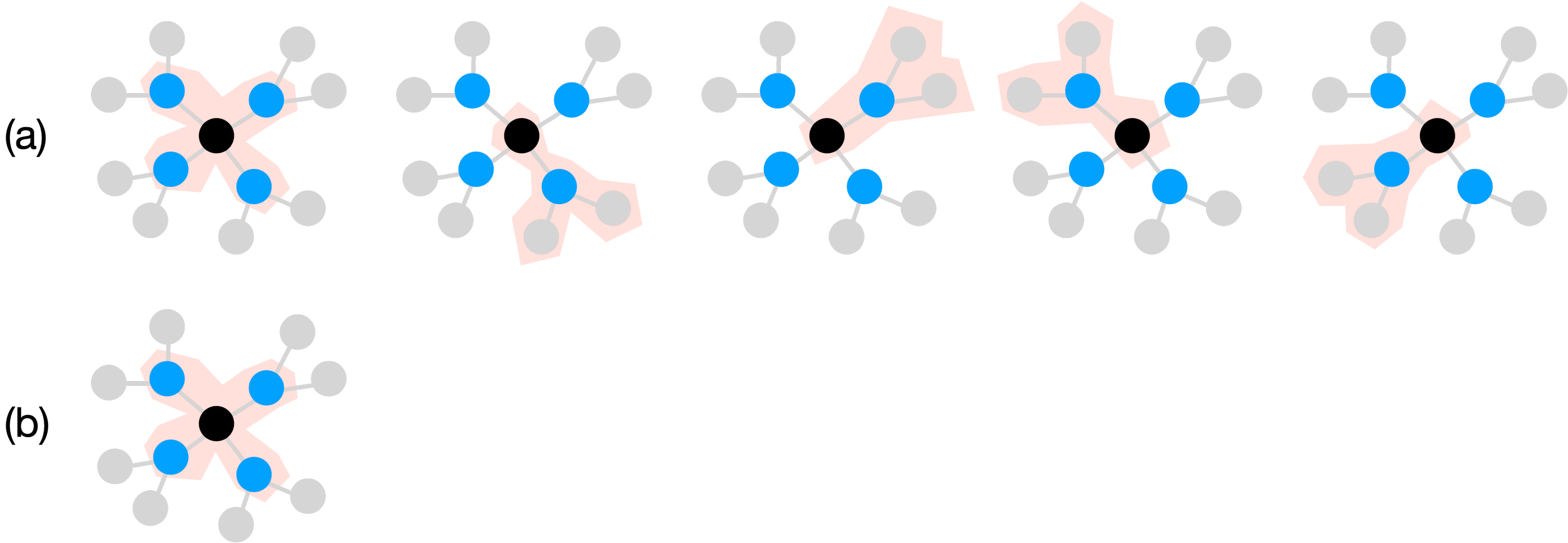}  
\end{center}   
\caption{Different definitions of the payoff in $N$-player games on networks.
(a) The definition of the payoff for the present STNG. A focal player (the black circle) belongs to five groups (shown as the shaded area in each of the five copies of the local network). The focal player is assumed to earn payoffs from each group.
The summed payoff drives evolutionary game dynamics, as is often the case for other $N$-player game dynamics on networks such as the PGG \cite{Santos:2008aa, Perc:2013aa}. 
(b) An alternative definition of the payoff in an $N$-player game on networks. With this definition, while
the focal player belongs to the same five groups, one assumes that the focal player's payoff
only originates from the one group centred around it. The network version of the original NTG uses this definition of the payoff \cite{Chica:2018wt}.
}
\label{fig_multiple_groups} 
\end{figure}

Our contributions include:
\begin{itemize}
    \item Proposal of an SNTG in which players alternate between the two roles, with the average payoff from the two roles driving evolutionary game dynamics.
    \item Analysis of evolutionary dynamics of the SNTG in the infinite well-mixed population and two types of finite structured populations (i.e., square lattice and heterogeneous networks).
    \item Examination of interactions between payoff function nonlinearity and population structure, affecting evolution of trust.
    \item Evidence that high-degree nodes influence evolution of trust in heterogeneous networks, suggesting potentials for intervention.
\end{itemize}
Principal differences between the original NTG and the proposed SNTG on networks are shown in Fig.\,\ref{fig_multiple_groups} and Table \ref{tab_comparison}. Further details on these distinctions are provided in subsequent sections.
 
\begin{table*}[t!]
\caption{Comparison of the original NTG and SNTG on networks. We denote by $d$ the degree of a focal player node.
Regarding the number of groups and players affecting a focal player's payoff, the SNTG is identical to the PGG for networks \cite{Santos:2008aa,Perc:2013aa}.
Because there is no network version of the asymmetric NTG proposed, we do not include the asymmetric NTG in the present comparison.}
\begin{center}
\begin{tabular}{|c|c|c|c|c|}
\hline
Game & \parbox[t]{4cm}{\centering{Number of groups affecting\\ the focal player's payoff}} & \parbox[t]{5cm}{\centering{Players affecting\\ the focal player's payoff}} & \parbox[t]{2.3cm}{\centering{Payoff obtained from}} & \parbox[t]{2.0cm}{\centering{Number of strategies}}\\
\hline
\hline
NTG \cite{Chica:2018wt} & 1 & neighbours & one role & 3\\
\hline
SNTG & $d+1$ & neighbours, neighbours of neighbours & both roles  & 4\\
\hline
\end{tabular}
\end{center}
\label{tab_comparison}
\end{table*}%

\section{Model}

\subsection{Strategies}

Our SNTG expands upon the asymmetric NTG \cite{Lim:2024ty} by symmetrising it.  Players alternate between investor and trustee roles, similar to symmetrising 2-player asymmetric games \cite{Lim:2022vt,Hofbauer:2003aa,sigmund2016calculus}.
Investors choose whether to invest or not, while trustees decide to be trustworthy or untrustworthy. 
A player's strategy therefore comprises two elements, one for each role: $i$ (i.e., invest) and $n$ (i.e., not to invest) as investor and $t$ (i.e., trustworthy) and $u$ (i.e., untrustworthy) as trustee. There are four possible strategies: $it, iu, nt$, and $nu$.

A total of $N_I$ investors and $N_T$ trustees participate in an NTG, where $N = N_I + N_T$. Each player among the $N_I$ investors employs their investor strategy in the NTG; for example, an $it$ player within the $N_I$ investors acts as an investing investor. Similarly, each player among the $N_T$ trustees utilises their trustee strategy; for instance, an $it$ player within the $N_T$ trustees acts as a trustworthy trustee. In an NTG, the profit from the investment made by investing investors is equally distributed among trustees, and then the trustworthy trustees share this profit with the investing investors \cite{Lim:2024ty}. 

\subsection{Payoffs}

Given $N_I$ investors and $N_T$ trustees in an NTG, the payoffs $\Pi_i$, $\Pi_n$, $\Pi_t$, and $\Pi_u$ for investing investors, non-investing investors, trustworthy trustees and untrustworthy trustees, respectively, are
\begin{equation}
\begin{aligned}
\Pi_i(m_i,g_t) &= \frac{g_t}{N_T} \frac{r\left(1-w^{m_i}\right)}{m_i(1-w)} +\frac{g_t}{N_T} -1, 
\quad \Pi_n = 0, \\
\Pi_t(m_i) &= r\frac{1}{N_T} \frac{1-w^{m_i}}{1-w}, 
\quad\quad\quad \Pi_u(m_i) = \frac{1}{r} \Pi_t(m_i),
\end{aligned}
\label{eq_payoffs_roles}
\end{equation}
where $m_i \in \{0,1,\ldots,N_I\}$ and $g_t \in \{0,1,\ldots,N_T\}$ are the numbers of investing investors and trustworthy trustees, respectively. 
Parameter $r$, satisfying $0 < r <1$, represents the productivity of prosocial strategies (i.e., investing investor and trustworthy trustee) relative to the payoff that an untrustworthy trustee receives from investing investors. Parameter $w$ $(>0)$ determines how the investment value accumulates.
For $0<w<1$, each additional investor's contribution diminishes. At $w=1$, each investor's contribution is $1$ regardless of $m_i$ (obtained using L'H\^{o}pital's
 rule). For $w>1$, the per-investor contribution increases with $m_i$, representing synergistic benefits.
See \cite{Lim:2024ty} for further details.

We examine the SNTG in both well-mixed and structured populations, where the expected payoff of each player across the two different roles drives the evolutionary game dynamics.

\subsection{Well-mixed Populations} 

In a well-mixed, infinitely large population, we denote by $y_{it}$, $y_{iu}$, $y_{nt}$, and $y_{nu}$ the frequencies of $it$, $iu$, $nt$, and $nu$ types, respectively, with $y_{it} +y_{iu} +y_{nt} +y_{nu}=1$. 
From time to time, $N_I$ investors and $N_T$ trustees are randomly selected to participate in a one-shot NTG. We define $N \equiv N_I + N_T$ and fix the values of $N_I$ and $N_T$. 

\subsubsection{Expected Payoffs}   

For an investing investor of $it$ or $iu$ type, the investor part of the expected payoff $P_i$ is    
\begin{multline}
P_i = -1 + (y_{it} + y_{nt}) \left[1 + \frac{r}{N_I(1-w)} \frac{1}{(y_{it} + y_{iu})} 
\left\{1 - \left[1 + (w-1)(y_{it} + y_{iu})\right]^{N_I}\right\}\right].
\label{eq_payoff_investing}
\end{multline}
Note that Eq.~\eqref{eq_payoff_investing} requires that $y_{it}+y_{iu}\ne0$ and $w\neq 1$. To ensure well-defined evolutionary dynamics across the entire state space of the simplex, $P_i$ must be defined everywhere.
For both $y_{it}+y_{iu}=0$ and $w=1$, we define $P_i$ using L'H\^{o}pital's rule.
See Appendix \ref{deri_payoff_investing} for the derivation.
For a non-investing investor of $nt$ or $nu$ type, the investor part of the expected payoff $P_n$ is
\begin{equation}
P_n =0.
\end{equation}
For a trustworthy trustee of $it$ or $nt$ type, the trustee part of the expected payoff $P_t$ is
\begin{equation}
\begin{split}
P_t
&=  r\frac{1}{N_T(1-w)} \left\{1 -\left[1 +(w-1) (y_{it}+y_{iu})\right]^{N_I}\right\}.
\end{split}
\label{eq_payoff_trustworthy}
\end{equation}
See Appendix \ref{deri_payoff_trustworthy} for the derivation.
For an untrustworthy trustee of $iu$ or $nu$ type, the trustee part of the expected payoff $P_u$ is
\begin{equation}
P_u =\frac{1}{r}P_t,
\label{eq_payoff_untrustworthy}
\end{equation}
which follows from $\Pi_u(k_{it}+k_{iu}) =\frac{1}{r}\Pi_t(k_{it}+k_{iu})$. 
The expected payoffs for players of $it$, $iu$, $nt$, and $nu$ types are
\begin{equation}
\begin{aligned}
P_{it} &= p_I P_i + (1-p_I)P_t, & P_{iu} &= p_I P_i + (1-p_I)P_u,\\
P_{nt} &= (1-p_I)P_t, & P_{nu} &= (1-p_I)P_u,
\label{eq_expected_payoffs}
\end{aligned}
\end{equation}
respectively, where $p_I=\frac{N_I}{N}\in\left\{\frac{1}{N},\ldots,\frac{N-1}{N}\right\}$ is the probability of each player taking an investor role. With the remaining probability $1-p_I$, each player takes a trustee role. Note that $N_I \in \{1,\ldots,N-1\}$ because an NTG requires at least one investor and one trustee; the game cannot be played without both roles present.

\subsubsection{Fermi Dynamics}

Occasionally, each player has the opportunity to imitate the strategy of another randomly chosen player. 
This imitation (also called social learning) process results in the following form of evolutionary game dynamics at the population level \cite{Sandholm:2010aa}:
\begin{equation}
\dot{y}_{s} =\sum_{s^{\prime} \in \mathcal{S}-\{s\}}y_{s^{\prime}} y_{s}\hat{\rho}_{s^{\prime} \rightarrow  s} -\sum_{s^{\prime} \in \mathcal{S}-\{s \}}y_{s}y_{s^\prime} \hat{\rho}_{s \rightarrow s^\prime},
\label{eq_dynamics}
\end{equation}
where the dot denotes the time derivative,
$s, s^\prime \in \mathcal{S}\equiv\{it,iu,nt,nu\}$,
and $\hat{\rho}_{s^\prime \rightarrow s}$ is proportional to the probability of a transition from strategy $s^\prime$ to $s$ through imitation.
The first term represents the inflow of players to strategy $s$ from the other strategies. The second term represents the outflow of players from strategy $s$ to the other strategies.
We employ the Fermi function for the strategy switching:
\begin{equation}
\hat{\rho}_{s_f\rightarrow s_m} \equiv\frac{1}{1+e^{-\beta  (P_m-P_f)}},
\label{eq_Fermi}
\end{equation}
where
$P_m$ and $P_f$ denote the expected payoffs of the mimicked player and mimicking player, respectively,
and $\beta \ge 0$ denotes the selection strength.
Equation~\eqref{eq_Fermi} is widely used in evolutionary game dynamics \cite{Szabo:2007tg,Perc:2013aa}.
Then, we obtain
\begin{equation}
\dot{y}_{s} =y_{s} \left[ \sum_{s^\prime \ne s} y_{s^{\prime}} \tanh \left(\frac{1}{2} \beta  (P_{s} -P_{s^\prime})\right)\right],
\label{eq_population_dynamics}
\end{equation}
where $s, s^\prime \in\mathcal{S} $, and
$\frac{1}{1+e^{-x}} -\frac{1}{1+e^x} =\tanh(\frac{x}{2})$ is used.
Note that, given the simplex constraint $y_{it}+y_{it}+y_{nt}+y_{nu}=1$, we are left with three independent variables. 
For weak selection, where $\beta \ll 1$, we have 
$\tanh \left(\frac{1}{2} \beta  (P_{s^\prime} -P_s)\right) \approx \frac{1}{2} \beta  (P_{s^\prime} -P_s)$ such that Eq.\,\eqref{eq_population_dynamics} reverts to replicator dynamics.

\subsection{Structured Populations}

In structured populations, individuals interact via edges of the given network. An SNTG is associated with a neighbourhood in the network, with each player usually belonging to multiple neighbourhoods. 
Given a player's node degree $d$, the player belongs to $d+1$ groups and thus participates in $d+1$ SNTGs: one associated with the player's own neighbourhood and those of the player's neighbours.
A player's total payoff in one round is assumed to be the sum of the payoffs from all $d+1$ groups. 
Notably, the strategies of neighbours of the neighbours of player $j$ affect $j$'s total payoff. This fact distinguishes $N$-player games from 2-player games on networks; in the latter case, only $j$'s direct neighbours impact $j$'s total payoff.
  
\subsubsection{Expected payoffs from a group}

From a group of $N = d+1$ players defined by a node with degree $d$ and its neighbours in a network, we select $N_I$ players as investors uniformly at random. The remaining $N_T$ $(= N - N_I)$ players act as trustees. These investors and trustees then participate in an NTG to earn payoffs. Both $N_I$ and $N_T$ are fixed for each group. 
In a structured population, the group size $N$ can vary, depending on the node degree. Therefore, we determine $N_I$ per group using a single global parameter $p \in (0,1)$, which is an approximate probability that a player takes an investor role. Given $p$, we set $N_I =\min\left(\lceil Np \rceil, N-1\right)$, yielding $N_I \in\{1,\ldots,N-1\}$. This ensures that the NTG remains well-defined, as previously described for the well-mixed population.

Due to the random selection of investors, the payoff for each strategy in this NTG is stochastic. Consequently, we focus on the expected payoffs, calculating one for each strategy in each group. Using these expected payoffs ensures
that the players employing the identical strategy gain the same payoff from the NTG played within the group.
The expected payoffs are given by:
\begin{align}
P_{i|it} &= -\frac{N_{iu}+N_{nu}}{N-1} 
+\frac{r}{(1-w)(N-N_I)\binom{N-1}{N_I-1}} 
\sum_{k_i=0}^{N_I-1} \binom{N_{it}+N_{iu}-1}{k_i}  \binom{N_{nt}+N_{nu}}{N_I -1 -k_i} \frac{\left(1-w^{k_i+1}\right)}{k_i+1}  \nonumber \\
& \quad \times\left[N_t-1 -(N_I -1)\Big\langle\frac{N_{nt}}{N_{nt}+N_{nu}}\Big\rangle_0 
+k_i\left(\Big\langle\frac{N_{nt}}{N_{nt}+N_{nu}}\Big\rangle_0 
-\Big\langle\frac{N_{it}-1}{N_{it}+N_{iu}-1}\Big\rangle_0\right)\right],
\label{eq_payoff_it_investor}
\\
P_{t|it}  &=\frac{r}{(1-w)(N-N_I)} 
\left[1 -\frac{1}{\binom{N-1}{N_I}} 
\sum_{k_i=0}^{N_I} \binom{N_{it}+N_{iu}-1}{k_i}
\binom{N_{nt}+N_{nu}}{N_I -k_i} w^{k_i}\right],
\label{eq_payoff_it_trustee}
\\
P_{i|iu}  &= -\frac{N_{iu}+N_{nu}-1}{N-1} 
+\frac{r}{(1-w)(N-N_I)\binom{N-1}{N_I-1}} 
 \sum_{k_i=0}^{N_I-1}  \binom{N_{it}+N_{iu}-1}{k_i}\binom{N_{nt}+N_{nu}}{N_I -1 -k_i} \frac{\left(1-w^{k_i+1}\right)}{k_i+1} \nonumber \\
& \quad \times \left[N_t -(N_I -1)\Big\langle\frac{N_{nt}}{N_{nt}+N_{nu}}\Big\rangle_0 
+k_i\left(\Big\langle\frac{N_{nt}}{N_{nt}+N_{nu}}\Big\rangle_0
-\Big\langle\frac{N_{it}}{N_{it}+N_{iu}-1}\Big\rangle_0\right)\right],
\label{eq_payoff_iu_investor}
\\
P_{u|iu} &= \frac{1}{r}P_{t|it},
\label{eq_payoff_iu_trustee}
\\
P_{n|nt} &= P_{n|nu} = 0,
\\
P_{t|nt}  &=\frac{r}{(1-w)(N-N_I)} \left[1 -\frac{1}{\binom{N-1}{N_I}} \sum_{k_i=0}^{N_I}  \binom{N_{it}+N_{iu}}{k_i}
\binom{N_{nt}+N_{nu} -1}{N_I -k_i} w^{k_i}\right],
\label{eq_payoff_nt_trustee}
\\
P_{u | nu}  &= \frac{1}{r} P_{t|nt},
\label{eq_payoff_nu_trustee}
\end{align}
where $P_{i|it}$, for example, represents the expected payoff for the $it$ player in the investing investor role;
$N_{it}, N_{iu}, N_{nt}$, and $N_{nu}$ represent the numbers of $it$, $iu$, $nt$ and $nu$ players, respectively, in the group,
$N=N_{it}+N_{iu}+N_{nt}+N_{nu}=N_I+N_T$,
and $\langle\frac{a}{b}\rangle_0 \equiv 
\begin{cases}
\frac{a}{b}, & b\ne 0\\
0, & b=0
\end{cases}
$.
See Appendix \ref{deri_eq_payoff_it_investor}, \ref{deri_eq_payoff_it_trustee}, and 
\ref{deri_eq_payoff_iu_nt_nu} for the derivation of Eqs.\,\eqref{eq_payoff_it_investor}--\eqref{eq_payoff_nu_trustee}. 
Note that we have $P_{i|it}\ne P_{i|iu}$, $P_{t|it} \ne P_{t|nt}$, and $P_{u | iu} \ne P_{u | nu}$. In other words, the expected payoff of a player as an investor may depend not only on its strategy as an investor but also on its strategy as a trustee, and vice versa. For instance, the payoff of a player as an investing investor differs depending on whether its trustee strategy is trustworthy or untrustworthy, $P_{i|it}\ne P_{i|iu}$. This contrasts with well-mixed populations, where the payoff of a player as an investor depends solely on its strategy as an investor, and similarly for the payoff as a trustee. This discrepancy between well-mixed and structured populations stems from differences in sampling. In a well-mixed population, investors and trustees are independently sampled from the entire infinite population using the multinomial distribution. In networks, investors are sampled from a group of $N$ players defined by a node and its neighbours, using the multivariate hypergeometric distribution.

For a given group, the expected payoffs $P_{it}, P_{iu}, P_{nt}$, and $P_{nu}$ for the $it$, $iu$, $nt$ and $nu$  players in it, respectively, are
\begin{equation}
\begin{aligned}
P_{it} &= p_IP_{i|it}+(1-p_I)P_{t|it}, &
P_{iu} &= p_I P_{i | iu}  +(1-p_I) P_{t | iu},\\
P_{nt} &=(1-p_I)P_{t|nt}, &
P_{nu} &=(1-p_I)P_{u|nu},
\end{aligned}
\end{equation}
where $p_I=\frac{N_I}{N}\in\left\{\frac{1}{N},\ldots,\frac{N-1}{N}\right\}$ is the probability of each player taking an investor role. With the remaining probability $1-p_I$, each player takes a trustee role.

\subsubsection{Simulations} 

We numerically run evolutionary dynamics on networks as follows. The simulation begins with an initial population of strategies, then repeatedly performs two steps:
(1) We uniformly randomly select a player (i.e., node), denoted by $j$, and one of its neighbours, denoted by $k$.
(2) Node $j$ adopts $k$'s strategy $s_k$  with a probability determined by the Fermi function of the payoff difference, Eq.~\eqref{eq_Fermi}.

Unless we state otherwise, simulations commence with equal proportions of the four strategies ($it$, $iu$, $nt$, and $nu$) uniformly randomly distributed over the $Z = 1024$ nodes, which we refer to as the unbiased initial condition.  Each simulation runs for 5000 generations.  In each generation, the two steps are repeated $Z$ times, allowing each node to update its strategy once per generation on average.
For each parameter pair $(w,r)$, we ran 50 independent simulations.

We use two networks. The first network is a square lattice of linear size $Z^{1/2} = 32$ with periodic boundary conditions and the von Neumann neighbourhood, which yields the degree of each node equal to $d=4$.  The second network is a realization of the Barab\'{a}si-Albert (BA) model, which we refer to as a heterogeneous network; the network construction begins with a cycle graph containing 3 nodes, i.e., a triangle, and then we add a node with $m=2$ edges in each step of the preferential attachment algorithm.
The average degree is equal to $\langle d\rangle \approx 3.994$, which is approximately the same as that for the square lattice by design.
We generate a different heterogeneous network for each simulation. The BA model produces an approximately power-law degree distribution with power-law exponent $3$, which is in stark contrast with the square lattice in which all nodes have the same degree.

\section{Results}

\subsection{Well-mixed Populations}

\begin{figure*}[t!]  
\begin{center}  
\includegraphics[width=0.94\textwidth]{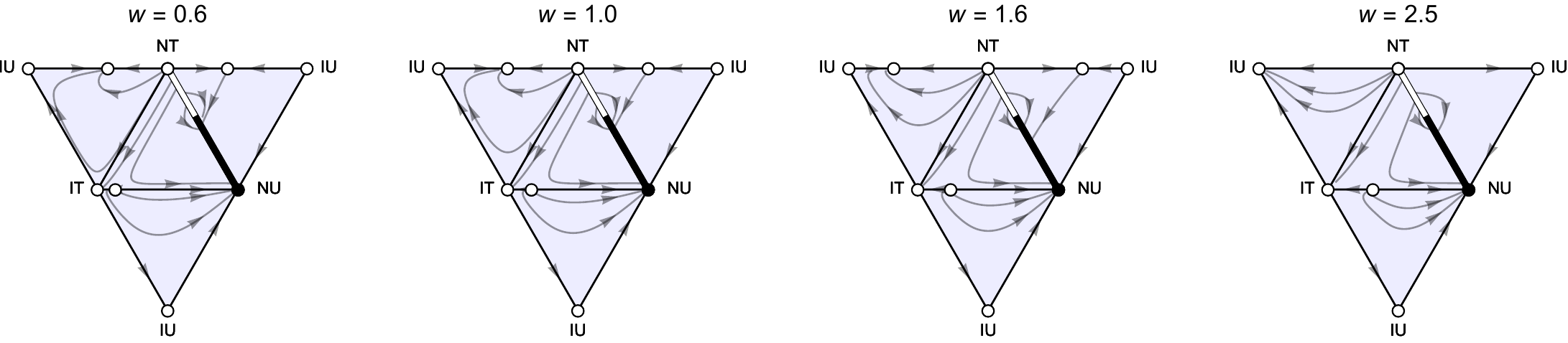}           
\end{center}    
\caption{
Evolutionary dynamics of the SNTG in the infinite well-mixed population, shown over the triangular faces of the 3-simplex $\triangle^3  = \{(y_{it},y_{iu},y_{nt},y_{nu}) : y_{it} +y_{iu} +y_{nt} +y_{nu}=1\}$ as the state space.
Vertices IT, IU, NT, and NU correspond to homogeneous population states $y_{it}=1$, $y_{iu}=1$, $y_{nt}=1$, and $y_{nu}=1$, respectively.
All trajectories converge to the line of stable equilibria on the NT-NU edge given by $y_{nt}+y_{nu}=1$ and $\frac{r}{r+1}< y_{nu}\le 1$.
Therefore, investment (i.e. trust) does not evolve.
Equilibria appear only on vertices and edges, but not in the interior of the faces.
A filled circle represents a stable equilibrium (i.e., NU). Open circles represent unstable equilibria (i.e., IT, IU, and NT).
On the NT-NU edge, the thick solid segment indicates stable equilibria; the hollow segment indicates unstable equilibria.
Nonlinearity in the payoff function (i.e., $w\ne 1$) does not qualitatively change evolutionary outcomes compared to linearity (i.e., $w=1$).
We set $N=5$, $N_I=3$, $r=0.8$, $\beta=10$, and $w\in\{0.6,1,1.6,2.5\}$.
}
\label{fig_well_mixed_simplex} 
\end{figure*}

In a well-mixed population, investment (i.e. trust) does not evolve (Figs.\,\ref{fig_well_mixed_simplex} and \ref{fig_well_mixed}). Consequently, the average payoff for a player is equal to $0$. All trajectories converge to the line of stable equilibria on the NT-NU edge, including the vertex NU (i.e., unanimity of $nu$ players), i.e., $y_{nt}+y_{nu}=1$ with $\frac{r}{r+1}< y_{nu}\le 1$,
of the 3-simplex $\triangle^3  = \{(y_{it},y_{iu},y_{nt},y_{nu}) : y_{it} +y_{iu} +y_{nt} +y_{nu}=1\}$ as the state space (Fig.\,\ref{fig_well_mixed_simplex}).

Vertices IT, IU and NT of the 3-simplex are unstable equilibria.
The entire interior of the IT-NU and IU-NT edges and part of interior of the NT-NU edge given by $0 < y_{nu} < \frac{r}{r+1}$ are also unstable
equilibria. There is no other equilibria including the interior of the triangles and the tetrahedron.
For proofs, see Appendix \ref{proof_vertices}, \ref{proof_edges},  \ref{proof_faces}, and \ref{proof_simplex}.      
The nonlinearity in the payoff, $w$, has no impact on the evolutionary dynamics.
Moreover, the evolutionary outcomes of the SNTG resemble those of the symmetric 2-player TG \cite{Lim:2022vt}, with both games ultimately resulting in $y_{nt} + y_{nu} = 1$, indicating no evolution of investment.

\subsection{Structured Populations}

Unlike in well-mixed populations, investment and trustworthiness can evolve on networks. Nonlinear payoff functions influence this evolution, either promoting or hindering it compared to linear payoff functions. Square lattices and heterogeneous networks produce different outcomes.

\subsubsection{Square Lattice}   

For the square lattice, we show the equilibrium fraction of each strategy and the average payoff over all the nodes 
in of Fig.\,\ref{fig_investor_probability}(a) as a function of $p$, $w$, and $r$. The figure indicates that sub-linearity ($w<1$) in payoff functions impedes evolution of $it$. Specifically, under sub-linearity, $it$ evolves in range of $r$ that is a proper subset of the range of $r$ under linearity ($w=1$).
This inhibitory effect intensifies as $w$ decreases or $p$ increases. 
Conversely, super-linearity ($w > 1$) facilitates the evolution of $it$, enabling it in a broader range of $r$ compared to linearity.
This effect tends to intensify as $p$ or $w$ increases.
Increasing $p$ is equivalent to increasing investor number $N_I$, with group size $N = N_I + N_T$ remaining constant. 
Note the non-monotonic behaviour in the evolution of $it$ for $w>1$, with a `trough' near $r=0.85$.
Figure\,\ref{fig_investor_probability}(a) suggests a threshold $r = r^*(w)$ such that $it$ evolves for $r > r^*$ for a given $w$. We observe that $r^*$ seems to decrease with $w$.
At $p=1/5$, there is no impact of $w$ since there is only one investor per group; impacts of $w$ require at least two investors in the group.

\subsubsection{Heterogeneous Network}

\begin{figure*}[t!]  
\begin{center}  
\includegraphics[width=0.94\textwidth]{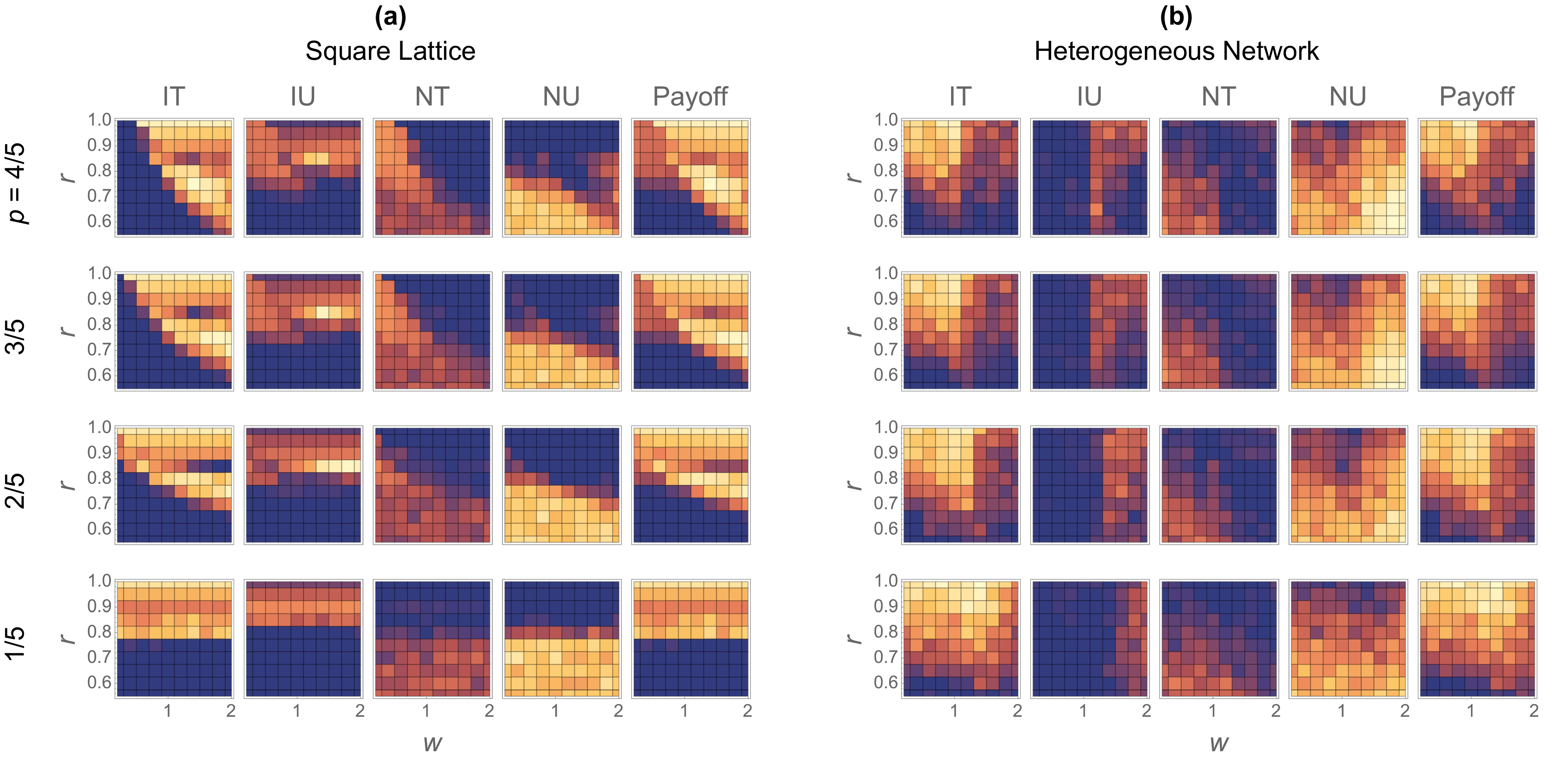}      
\includegraphics[width=0.23\textwidth]{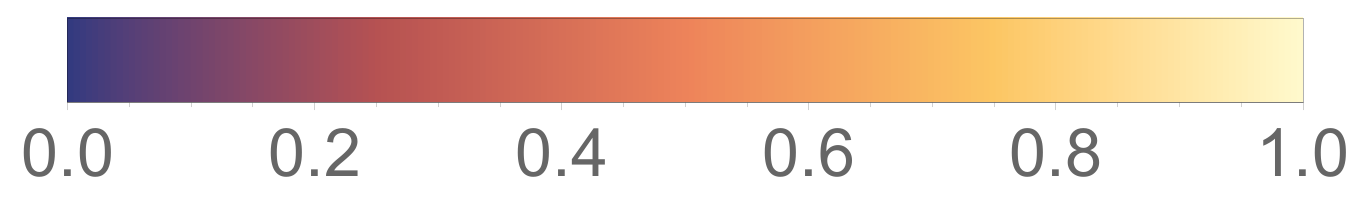}  
\end{center}   
\caption{Properties of approximate equilibria of the SNTG in finite networks.
(a) Square lattice. (b) Heterogeneous networks. In both (a) and (b), the first four columns 
show the fraction of each strategy as the function of $w$ and $r$. The fifth column shows 
the average payoff over all nodes, normalised by that of a population comprised entirely of $it$.
We used $(w,r) \in \left\{
0.2,0.4,0.6,0.8,1,1.2,1.4,1.6,1.8,2\right\} \otimes \left\{
0.55, 0.6, 0.65, 0.7, 0.75, 0.8,0.85, 0.9, 0.95, 1\right\}$.
We ran 50 simulations per parameter set and for 5000 generations each.
For each $(w, r)$ pair, we obtained the approximate equilibrium fraction of each strategist as the average over the last 256 generations for each simulation and over the 50 simulations.
We used $p \in \{ 1/5, 2/5, 3/5, 4/5 \}$.
}
\label{fig_investor_probability} 
\end{figure*}

We show the results for the heterogeneous networks in Fig.\,\ref{fig_investor_probability}(b).
We observe that heterogeneous networks produce markedly different outcomes compared to square lattices.
Under sub-linearity ($w<1$), heterogeneous networks promote the evolution of $it$ less than under linearity, but this inhibitory effect is less pronounced than in the square lattice.
Under super-linearity ($w>1$), the inhibitory effect is stronger than under sub-linearity and intensifies as $p$ or $w$ increases. This result sharply contrasts with that on the square lattice, for which super-linearity acts as a catalyst for the evolution of $it$ in a broader $r$ range than linearity does.
At a low probability of $p =1/5$, heterogeneous networks facilitate $it$ in a wider $r$ range than the square lattice, regardless of the $w$ value. For a larger $p$ value, this advantage of heterogeneous networks is limited to a narrower range of $w$, e.g., $w < 1$.

\subsubsection{Analytical Insights} 

\begin{figure}[t!]     
\begin{center}     
\includegraphics[width=0.49\textwidth]{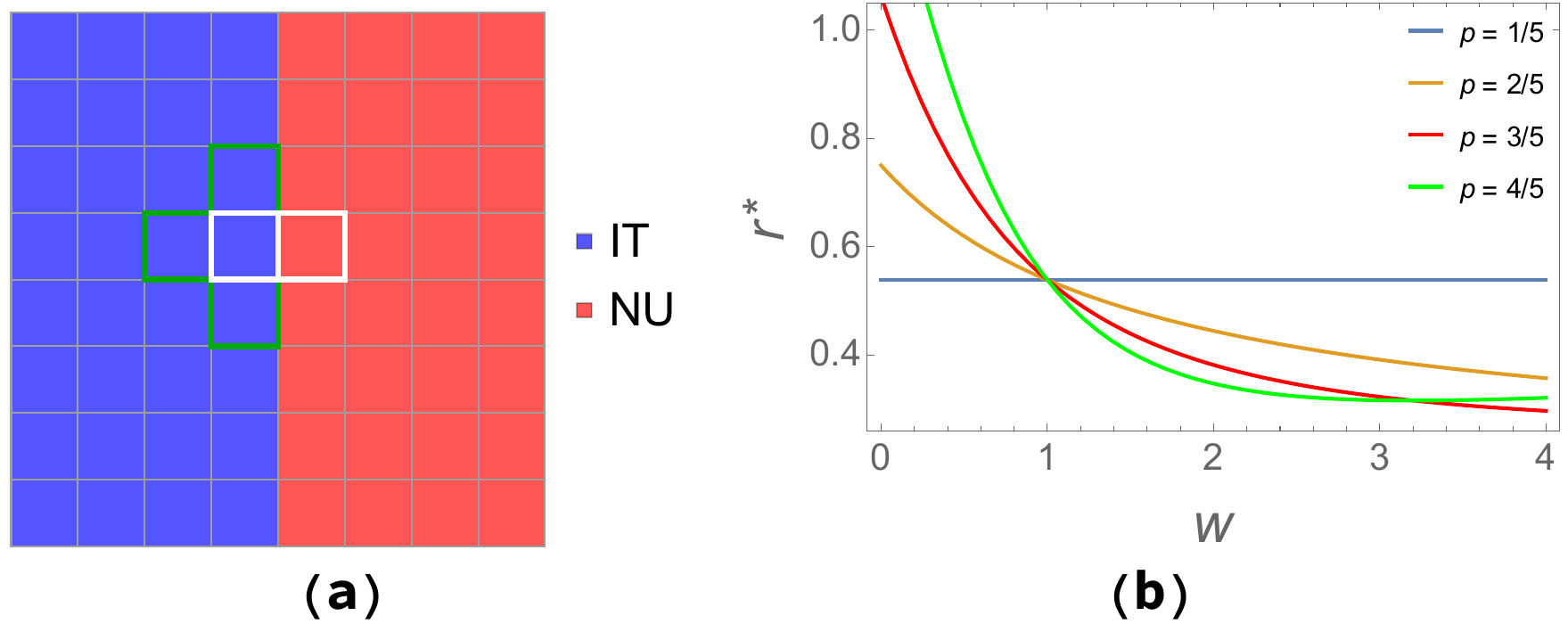}  
\end{center}   
\caption{Analytical approximation to the threshold $r^*$ for the square lattice.
One obtains $P_{it} = P_{nu}$ at $r=r^*$. The threshold $r^*$ is derived from a configuration composed of an $it$ cluster and an $nu$ cluster on the infinite square lattice shown in (a). 
Under this configuration, the strategy can only change on the border between the two clusters, as shown by the two square cells with white boundaries. An $it$ player may become $nu$ by imitating its $nu$ neighbour or vice versa, depending on whether
$P_{it} < P_{nu}$ (corresponding to $r <r^*$) and $P_{it} > P_{nu}$ (corresponding to $r>r^*$), respectively.}
\label{fig_analysis_IT_NU_homogeneous}  
\end{figure}

To gain analytical insights into the simulation results for the square lattice, we examine a simple configuration on the infinite square lattice. We initialise the grid with $nu$ players on one half and $it$ players on the other half, creating a straight border between the $nu$ players and $it$ players (Fig.\,\ref{fig_analysis_IT_NU_homogeneous}(a)).
Strategy switching occurs only on the border. 
If and only if an $nu$ player's payoff on the border is lower than that of an $it$ neighbour (i.e., $P_{nu} < P_{it}$), the region of $it$ is likely to invade the region of $nu$ over time.
For a given $w$, there exists a unique threshold $r^*$ such that $P_{it} > P_{nu}$ if and only if $r > r^*$. We obtain $r^*$ as the unique solution of the equation $P_{it} - P_{nu} = 0$.
We derive in Appendix \ref{deri_threshold_r_homo} that
$r^*=\frac{7}{13}$, $r^*=\frac{w+6}{5 w+8}$, $r^*=\frac{2 w^2+2 w+17}{7 w^2+16 w+16}$, and $r^*=\frac{w^3+w^2+w+11}{2 \left(w^3+4 w^2+4 w+4\right)}$ for $p = 1/5, 2/5, 3/5$, and $4/5$, respectively.
In Fig.\,\ref{fig_analysis_IT_NU_homogeneous}(b), we show $r^*$ as a function of $w$ for these four $p$ values.
The figure indicates that, for $p = 2/5, 3/5$ and $4/5$, the threshold $r^*$ strictly decreases as $w$ increases, up to $w=2$, demonstrating inhibition of $it$ when $w<1$ and promotion when $w>1$. Higher $p$ values yield a stronger dependence of $r^*$ on $w$. For $p = 1/5$, the threshold $r^*$ is independent of $w$.
This is trivial since $p = 1/5$ implies only one investor, whereas the payoff nonlinearity, $w$, has effects only when there are at least two investors in a group. 
These trends qualitatively align well with simulation results on the square lattice shown in Fig.\,\ref{fig_investor_probability}(a).

To obtain analytical insights into the simulation results for heterogeneous networks, we examine a double-star configuration \cite{Santos:2008aa}. 
We place a star graph with $N^{(1)}-1$ $nu$ players and another star graph with $N^{(2)}-1$ $it$ players. 
The $nu$ hub has degree $N^{(1)}-1$. The $it$ hub has degree $N^{(2)}-1$. 
The $nu$ and $it$ hubs are adjacent to each other. Each leaf node has degree $1$ and participates in a 2-player TG, which enforces $p_I =1/2$ for each group composed of a leaf node and its hub neighbour. 
As we did for the square lattice, we seek the threshold $r^*$ such that $P_{it}|\text{hub} > P_{nu}|\text{hub}$ if and only if $r >r^*$, where ``$|\text{hub}$'' indicates the expected payoff for the hubs.
This condition prevents the $nu$ hub from invading the $it$ hub. 
Solving $P_{it}|\text{hub} = P_{nu}|\text{hub}$ for $r$ yields
\begin{equation}
r^* = \frac{\text{num}}{\text{denom}},
\label{eq_threshold_hetero}
\end{equation}  
where
$\text{num} \equiv w [N^{(1)} \left(N^{(2)}\right)^2 p w - N^{(1)} \left(N^{(2)}\right)^2 p + (N^{(1)}-1) (N^{(2)}-1) w^{N^{(2)} p} - N^{(1)} N^{(2)} + N^{(1)} - N^{(2)} p w + N^{(2)} p + N^{(2)} - 1]$ and
$\text{denom} \equiv (N^{(1)}-1) \{(N^{(2)}-1) w [(N^{(2)}-2) N^{(2)} (w-1) - 2] - 2 w^{N^{(2)} p} [N^{(2)} (p-1) w - N^{(2)} p + w]\}$.
In the limit $N^{(1)}, N^{(2)} \rightarrow \infty$, Eq.\,\eqref{eq_threshold_hetero} simplifies to
\begin{equation}
r^*_{\infty} =
\begin{cases}
0, & 0 < w \le 1,\\
\frac{w}{2[p +(1 -p)w]}, & w > 1.
\end{cases}
\label{eq_threshold_hetero_infinity}
\end{equation}
For derivations of Eqs.\,\eqref{eq_threshold_hetero} and \eqref{eq_threshold_hetero_infinity}, see Appendix \ref{deri_threshold_r_hetero}.

\begin{figure*}[t!]   
\begin{center}      
\includegraphics[width=1\textwidth]{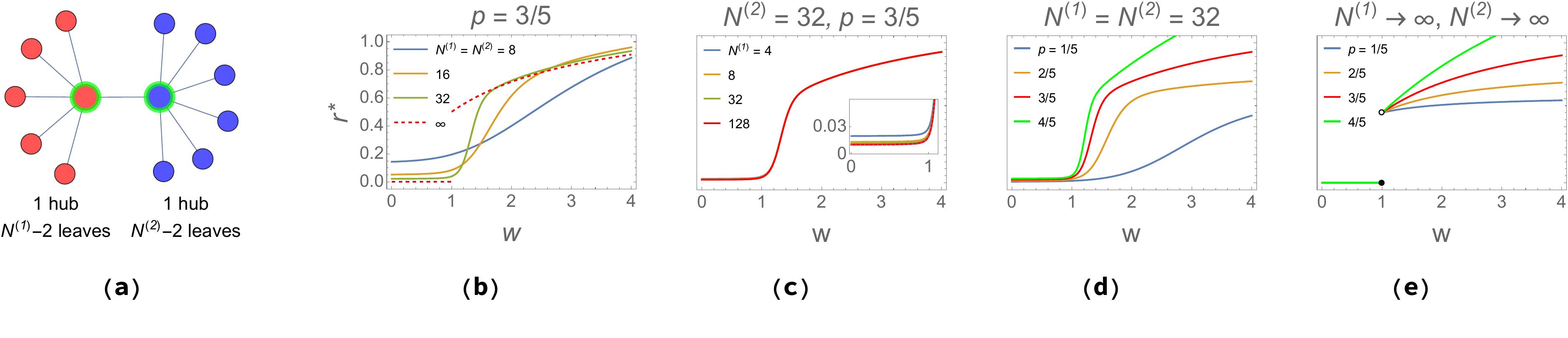}               
\end{center}   
\caption{Analytical approximation to $r^*$ from heterogeneous networks.
We consider two interconnected stars composed of $N^{(1)}-1$ nodes with strategy $nu$ and $N^{(2)}-1$ nodes with strategy $it$, respectively, as shown in (a).
Under super-linearity $w>1$, $r^*$ increases strictly with $w$.
 The difference between $N^{(1)}$ and $N^{(2)}$ has minimal impact, as shown in (c) and its inset.
A higher probability $p$ of being an investor leads to a steeper increase in $r^*$ for $w>1$, as shown in (d) and (e).
Larger group sizes $N^{(1)}$ and $N^{(2)}$ result in a more rapid increase in $r^*$ near $w=1$, as shown in (b) and (e).
These behaviours of $r^*$ are consistent with the numerical results shown in Fig.\,\ref{fig_investor_probability}(b). 
}
\label{fig_analysis_IT_NU_heterogeneous} 
\end{figure*}

For $w > 1$, $r^*$ increases with $w$, indicating stronger hindrance to the evolution of $it$ in heterogeneous networks with higher $w$ (Fig.\,\ref{fig_analysis_IT_NU_heterogeneous}(b) to (e)). For $w > 1$, $r^*$ also increases with $p$, indicating stronger hindrance with higher $p$ (Fig.\,\ref{fig_analysis_IT_NU_heterogeneous}(d) and (e)). 
For $w<1$, there is little difference in $r^*$ compared to that of $w=1$ (Fig.\,\ref{fig_analysis_IT_NU_heterogeneous}(b) to (e)). 
These trends qualitatively align well with the simulation results on heterogeneous networks shown in Fig.\,\ref{fig_investor_probability}(b). 
When hub nodes have lower degrees, $it$ tends to evolve more easily (i.e., $it$ evolves in a wider range of $r$ due to $r^*$ being lowered) for $w>1$ and less easily for $w<1$ (Fig.\,\ref{fig_analysis_IT_NU_heterogeneous}(b)). However, the disparity between the degree of the two hub nodes has a negligible impact (Fig.\,\ref{fig_analysis_IT_NU_heterogeneous}(c)). 

We remark the stark contrast between these analytical insights for the square lattice and heterogeneous networks.
On the square lattice, $it$ evolves more easily for $w>1$ than for $w=1$, whereas the opposite is the case on heterogeneous networks. In addition, on the square lattice, $it$ evolves less easily for $w<1$ than for $w=1$, whereas such a dependency on $w$ is absent on heterogeneous networks.

\subsubsection{Degree-Based Initialisation}

\begin{figure}[t!]  
\begin{center} 
\includegraphics[width=0.5\textwidth]{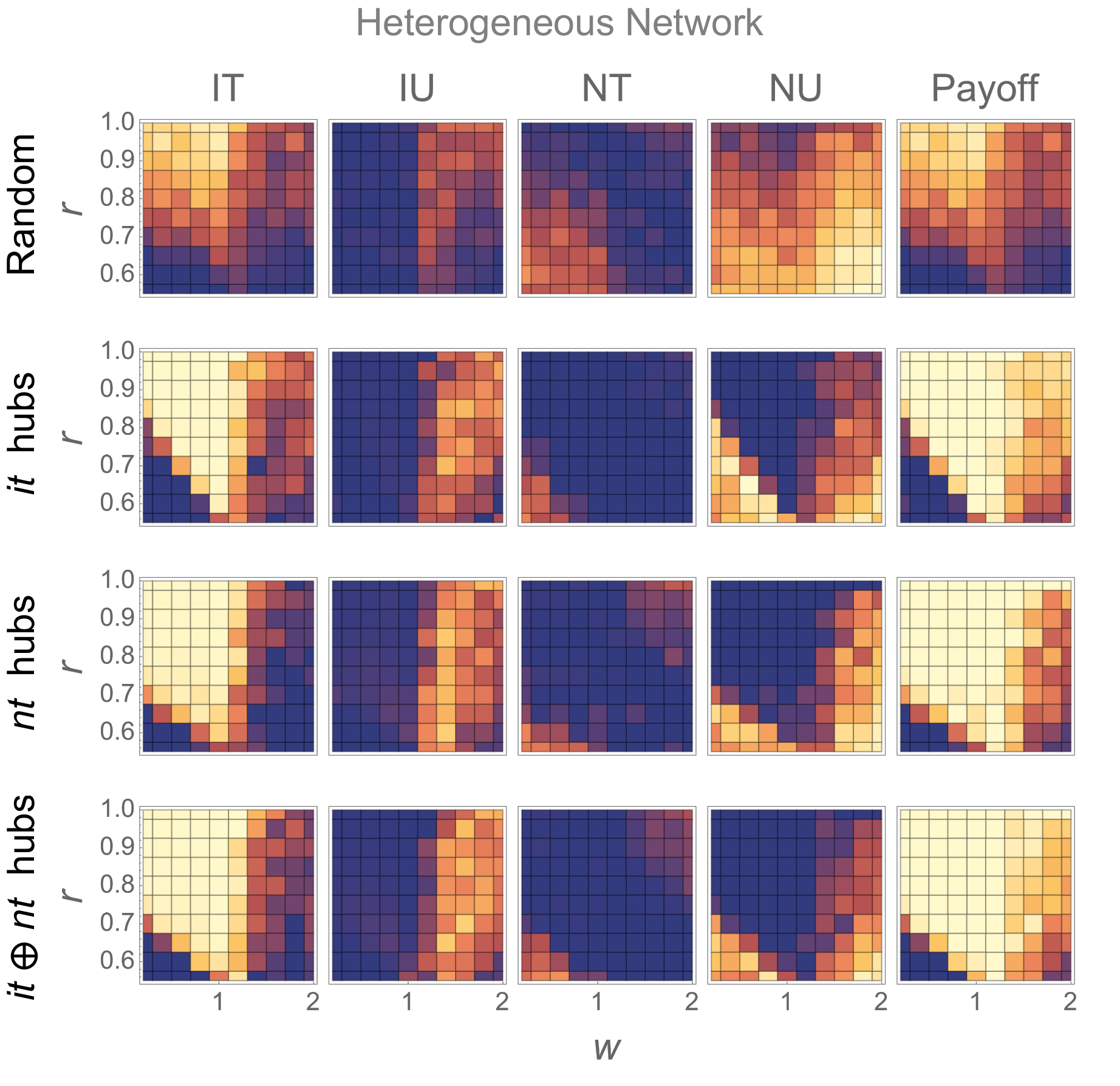} 
\\
\includegraphics[width=0.23\textwidth]{symmetric_NTG_Fig_unit_32}  
\end{center}   
\caption{Results for the degree-based initialisation in heterogeneous networks. 
`Random' refers to the case where all strategies are initially allocated to nodes uniformly at random, irrespective of the node's degree, as is the case for $p=3/5$ in Fig.\,\ref{fig_investor_probability}(b).
`$it$ hubs' refers to the initial condition in which the largest-degree nodes are inhabited by $it$, and the other three strategies are assigned to lower-degree nodes uniformly at random. 
Similarly, `$nt$ hubs' refers to the initial condition in which the largest-degree nodes are inhabited by $nt$. `$it\oplus nt$ hubs' refers to the initial condition in which the largest-degree nodes are inhabited by either $it$ or $nt$.
See the caption of Fig.\,\ref{fig_investor_probability} for how to read this figure.
}
\label{fig_hubs} 
\end{figure}

Here we investigate the impacts of degree-based initialization in heterogeneous networks, assuming the equal initial fraction of the four strategies (i.e., 25\% each).
We initiate $it$ at hubs (i.e., the top 25\% of nodes by degree), with other nodes uniformly randomly assigned $iu$, $nt$, and $nu$. 
This initial condition considerably increases the final fraction of $it$ and decreases those  $nt$ and $nu$ across a wider range of parameters than with the degree-independent random initialisation (see the second row of Fig.\,\ref{fig_hubs}).
Intriguingly, initiating $nt$ at hubs does not enhance $nt$ but promotes $it$ to a similar extent as initiating $it$ at hubs (see the third row of Fig.\,\ref{fig_hubs}). Initialization of the hubs by $nt$ fosters $it$ evolution more effectively at low $w$ and $r$ than the initialization of the hubs by $it$ does, and vice versa. To our surprise, a combination of the two (i.e. randomly initiating $it$ or $nt$ at hubs) promotes evolution of $it$ more effectively than either approach alone across the full range of parameters (see the fourth row of Fig.\,\ref{fig_hubs}).

\subsection{Robustness}  

\begin{figure*}[t!]  
\begin{center}  
\includegraphics[width=0.94\textwidth]{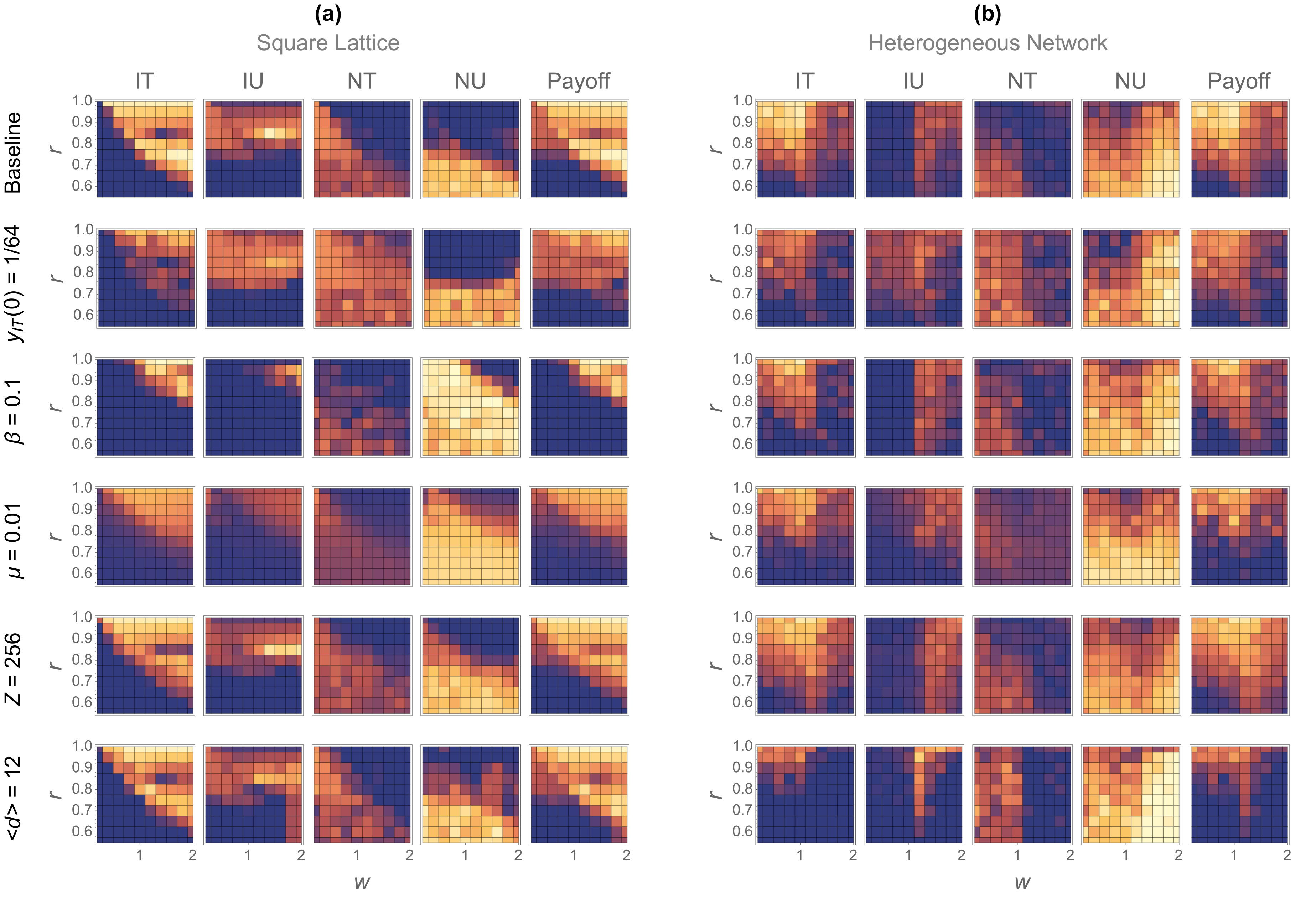}          
\includegraphics[width=0.23\textwidth]{symmetric_NTG_Fig_unit_32}  
\end{center}   
\caption{ 
Robustness of evolutionary outcomes.
The baseline case, shown in Fig.\,\ref{fig_investor_probability}, uses $p=3/5$, $y_{it}(0)=y_{iu}(0)=y_{nt}(0)=y_{nu}(0)=1/4$, $\beta =10$, $\mu =0$, and $Z =1024$.
In the second row of the figure panels, we reduced the initial frequency of $it$ to $y_{it}(0) =1/64$ from $1/4$, while maintaining $y_{iu}(0)=y_{nt}(0)=y_{nu}(0)=\left(1-y_{it}(0)\right)/3$.
In the third row, we decreased the selection strength to $\beta =0.1$ from $10$.
In the fourth row, we increased the mutation rate to $\mu =0.01$ from $0$.
In the fifth row, we decreased the population size to $Z =256$ from $1024$.
In the sixth row, we increased the mean node degree to $\langle d\rangle =12$ from $4$.
For additional results demonstrating robustness, see Figures\,\ref{fig_initial_condition}, \ref{fig_selection_strength}, \ref{fig_mutation_rate}, \ref{fig_population_size}, and \ref{fig_node_degree}. 
} 
\label{fig_robustness} 
\end{figure*}    

The equilibrium of the evolutionary dynamics is robust against variations in parameter values.
We demonstrated that order-of-magnitude changes in initial conditions, selection strength, mutation rate, population size or mean node degree produce outcomes qualitatively similar to those of the baseline (see Fig.\,\ref{fig_robustness}).
Notably, the key distinctions in outcomes between the square lattice and heterogeneous networks are well preserved. 

\section{Discussion}   


We propose the SNTG and analyse its evolutionary game dynamics. Each player is assumed to play both investor and trustee roles in each generation. Therefore, the mean payoff from both roles drives evolutionary dynamics, unlike the asymmetric NTG in which players have fixed roles (i.e., investor or trustee) and therefore different strategy sets over the evolutionary dynamics \cite{Lim:2024ty}. 
The SNTG reflects scenarios of role-switching. 
This approach aligns with behavioural experiments of TGs, 
as well as real-world interactions (such as buyer--seller exchanges), 
in which participants alternate roles \cite{Johnson:2011aa}.
The original NTG is technically a symmetric game because all players share the same set of strategies \cite{Abbass:2016aa}. In that game, each player can switch the role as a result of payoff-driven imitation (e.g., an investor turns into an untrustworthy trustee). The original NTG, in which each player has a fixed role at any given time and single-role payoffs drive evolutionary game dynamics, fundamentally differs from the present SNTG, in which each player plays both roles in any generation and the sum of the payoffs from the two roles drives evolutionary dynamics.


We have found that the SNTG is a challenging symmetric $N$-player game for prosocial behaviour to evolve.  The PGG is also a challenging, and widely studied, symmetric $N$-player games for evolution of prosocial behaviour. However, it can still foster the evolution of prosocial behaviour in well-mixed populations using nonlinear payoff functions \cite{hauert2006synergy,archetti2012review,Hauert:2024qy}. In contrast, we have shown that the SNTG does not foster prosocial behaviour even with nonlinear payoffs. 
The asymmetric NTG has been shown to present greater challenges for the evolution of prosocial behaviours than the PGG \cite{Lim:2024ty}. In-depth comparisons of the SNTG and PGG may be a productive exercise because both are symmetric games.


The interaction between nonlinearity and population structure differs between the SNTG and PGG. In both games, population structure alone promotes prosocial behaviours, even with linear payoffs.
In the SNTG, nonlinearity catalyses the evolution of prosocial behaviours with network reciprocity but does not promote prosocial behaviours on its own. In the PGG, nonlinearity directly boosts prosocial behaviours \cite{li2016evolutionary}. Here, the interaction is between two boosting mechanisms. Few other catalysts exist in $N$-player game evolutionary dynamics, apart from voluntary PGG participation \cite{Hauert:2007fk,Szabo:2002aa}.
A majority of research comparing homogeneous and heterogeneous networks in social dilemma games focuses on linear payoff functions, typically using 2-player or linear $N$-player games \cite{Chica:2018wt,Santos:2008aa,Santos:2005ly,Santos:2006aa,Perc:2013aa}. While some studies have examined nonlinear PGG in structured populations, they often focus on regular graphs \cite{li2016evolutionary,Szolnoki:2010aa}.
In the present study, we have examined both linear and nonlinear payoffs combined with two qualitatively different networks. Exploring the effects of nonlinear payoffs in $N$-player games may be worthwhile \cite{archetti2012review}, including the case of
various realistic networks.


We have also shown that, in heterogeneous networks, initialising based on the node degree can significantly enhance trust, offering the potential for effective interventions to promote prosocial behaviours. This approach requires only one-off involvement (e.g. initial incentives for hubs to engage in prosocial behaviours). After this initial intervention, all nodes alter their behaviours only through payoff-driven imitation, contrasting with schemes requiring continuous monitoring and intervention \cite{Sasaki:2012rz, Lim:2024ty}.
In the PD, the effect of degree-dependent initialisation is not univocal. Corroborating to our results, initialising cooperation at hubs can make the evolution of cooperation easier under imitation-based strategy updating \cite{Chen:2008aa}. However, degree-dependent initialisation only has moderate effects when updating is based on rational decision-making \cite{Dercole:2019aa}. These findings may trigger further study of effects of hubs through investigations of, for example, alternative strategy updating rules and different games.


We have proposed a symmetric $N$-player Trust Game and analysed its evolutionary dynamics. We find that prosocial behaviour 
can evolve only in structured populations (i.e., networks), in which sense maintaining prosocial behaviour is more challenging in this game compared to other symmetric $N$-player games. Nonlinear payoffs and network structure significantly influence evolution of prosocial behaviour, with the square lattice and heterogeneous networks yielding strikingly different outcomes. These results highlight the complex interplay between payoff structures and network structure in shaping prosocial behaviours in multi-agent systems.

\appendix
\section{Appendix}
\renewcommand{\theequation}{A.\arabic{equation}}	
\renewcommand{\thefigure}{A.\arabic{figure}}

\subsection{Derivation of Eq.\,\eqref{eq_payoff_investing}
\label{deri_payoff_investing}}

For an investor in a group of $N$ players with $N_I$ investors and $N_T$ trustees, the probability of having $k_{it}$, $k_{iu}$, $k_{nt}$, and $k_{nu}$ co-players of $it$, $iu$, $nt$, and $nu$ types among the remaining $N_I-1$ investors obeys the multinomial distribution given by
\begin{equation}
\Pr(k_{it},k_{iu},k_{nt},k_{nu}; N_I-1) =\frac{(N_I-1)!}{k_{it}!k_{iu}!k_{nt}!k_{nu}!} y_{it}^{k_{it}}y_{iu}^{k_{iu}}y_{nt}^{k_{nt}}y_{nu}^{k_{nu}}.
\end{equation}
Here, $y_{it}$, $y_{iu}$, $y_{nt}$, and $y_{nu}$ are the fraction of each type in the population, and $k_{it}+k_{iu}+k_{nt}+k_{nu}=N_I-1$.
The probability of having $l_{it}$, $l_{iu}$, $l_{nt}$, and $l_{nu}$ of each type 
among $N_T$ trustees is
\begin{equation}
\Pr(l_{it},l_{iu},l_{nt},l_{nu}; N_T) =\frac{N_T!}{l_{it}!l_{iu}!l_{nt}!l_{nu}!} y_{it}^{l_{it}}y_{iu}^{l_{iu}}y_{nt}^{l_{nt}}y_{nu}^{l_{nu}},
\end{equation}
where $l_{it}+l_{iu}+l_{nt}+l_{nu}=N_T$.
For an investing investor (i.e., $it$ or $iu$), the total number of investing investors is $k_{it}+k_{iu}+1$, and that of trusting trustees is $l_{it}+l_{nt}$.
Thus, the expected payoff $P_{i}$ for a player acting as an investing investor is given by
\begin{subequations}
\begin{align}
P_{i} &= \sum_{k_{it}+k_{iu}+k_{nt}+k_{nu}=N_I-1} \Pr(k_{it},k_{iu},k_{nt},k_{nu}; N_I-1) \sum_{l_{it}+l_{iu}+l_{nt}+l_{nu}=N_T} \Pr(l_{it},l_{iu},l_{nt},l_{nu}; N_T) \Pi_i(k_{it}+k_{iu}+1, l_{it}+l_{nt}) \notag \\
&= \sum_{k_{it}+k_{iu}+k_{nt}+k_{nu}=N_I-1} \frac{(N_I-1)!}{k_{it}!k_{iu}!k_{nt}!k_{nu}!} y_{it}^{k_{it}}y_{iu}^{k_{iu}}y_{nt}^{k_{nt}}y_{nu}^{k_{nu}} \sum_{l_{it}+l_{iu}+l_{nt}+l_{nu}=N_T} \frac{N_T!}{l_{it}!l_{iu}!l_{nt}!l_{nu}!} y_{it}^{l_{it}}y_{iu}^{l_{iu}}y_{nt}^{l_{nt}}y_{nu}^{l_{nu}} \notag \\
&\quad \times \left[\frac{l_{it}+l_{nt}}{N_T} \frac{r(1-w^{k_{it}+k_{iu}+1})}{(k_{it}+k_{iu}+1)(1-w)} -\left(1 -\frac{l_{it}+l_{nt}}{N_T}\right)\right] \notag \\
&= -1 +\sum_{k_{it}+k_{iu}+k_{nt}+k_{nu}=N_I-1} \frac{(N_I-1)!}{k_{it}!k_{iu}!k_{nt}!k_{nu}!} y_{it}^{k_{it}}y_{iu}^{k_{iu}}y_{nt}^{k_{nt}}y_{nu}^{k_{nu}} \left[\frac{r(1-w^{k_{it}+k_{iu}+1})}{(k_{it}+k_{iu}+1)(1-w)} +1\right] 
\notag \\
&\quad \times  
\sum_{l_{it}+l_{iu}+l_{nt}+l_{nu}=N_T} \frac{N_T!}{l_{it}!l_{iu}!l_{nt}!l_{nu}!} y_{it}^{l_{it}}y_{iu}^{l_{iu}}y_{nt}^{l_{nt}}y_{nu}^{l_{nu}} \frac{l_{it}+l_{nt}}{N_T} \notag \\
&= -1 +\sum_{k_{it}+k_{iu}+k_{nt}+k_{nu}=N_I-1} \frac{(N_I-1)!}{k_{it}!k_{iu}!k_{nt}!k_{nu}!} y_{it}^{k_{it}}y_{iu}^{k_{iu}}y_{nt}^{k_{nt}}y_{nu}^{k_{nu}} \left[\frac{r(1-w^{k_{it}+k_{iu}+1})}{(k_{it}+k_{iu}+1)(1-w)} +1\right] \frac{N_T y_{it} +N_T y_{nt}}{N_T}   \label{eq_pi_multinomial} \\
&= -1 +(y_{it} +y_{nt}) \left[1 +\frac{r}{1-w}\sum_{k_i=0}^{N_I-1} \sum_{k_{it}+k_{iu}=k_i} \sum_{k_{nt}+k_{nu}=N_I-1-k_i} \frac{(N_I-1)!}{k_{it}!k_{iu}!k_{nt}!k_{nu}!} y_{it}^{k_{it}}y_{iu}^{k_{iu}}y_{nt}^{k_{nt}}y_{nu}^{k_{nu}} \frac{1-w^{k_{it}+k_{iu}+1}}{k_{it}+k_{iu}+1} \right] \notag \\
&= -1 +(y_{it} +y_{nt}) \left[1 +\frac{r}{1-w}\sum_{k_i=0}^{N_I-1}(N_I-1)! \frac{1-w^{k_i+1}}{k_i+1} \sum_{k_{it}+k_{iu}=k_i} \frac{1}{k_{it}!k_{iu}!} y_{it}^{k_{it}}y_{iu}^{k_{iu}} \sum_{k_{nt}+k_{nu}=N_I-1-k_i} \frac{1}{k_{nt}!k_{nu}!} y_{nt}^{k_{nt}}y_{nu}^{k_{nu}} \right] \notag \\
&= -1 +(y_{it} +y_{nt}) \left[1 +\frac{r}{1-w}\sum_{k_i=0}^{N_I-1} \frac{1}{k_i+1} \frac{(N_I-1)!}{k_i!(N_I-1-k_i)!} (1-w^{k_i+1}) 
\sum_{k_{it}+k_{iu}=k_i} \frac{k_i!}{k_{it}!k_{iu}!} y_{it}^{k_{it}}y_{iu}^{k_{iu}} 
\right. \notag \\
&\quad \left. \times 
\sum_{k_{nt}+k_{nu}=N_I-1-k_i} \frac{(N_I-1-k_i)!}{k_{nt}!k_{nu}!} y_{nt}^{k_{nt}}y_{nu}^{k_{nu}} \right] \notag \\
&= -1 +(y_{it} +y_{nt}) \left[1 +\frac{r}{1-w}\sum_{k_i=0}^{N_I-1} \frac{1}{k_i+1} \binom{N_I-1}{k_i} (1-w^{k_i+1}) (y_{it}+y_{iu})^{k_i} (y_{nt} +y_{nu})^{N_I-1-k_i} \right] \label{eq_pi_binomial}  \\
&= -1 +(y_{it}+y_{nt}) \left[1 +\frac{r}{1-w}\sum_{k_i=0}^{N_I-1} \frac{1}{N_I} \binom{N_I}{k_i+1} (y_{it}+y_{iu})^{k_i} (1 -y_{it}-y_{iu})^{N_I-1-k_i} (1-w^{k_i+1}) \right] \notag \\
&= -1 +(y_{it}+y_{nt}) \left[1 +\frac{r}{N_I(1-w)(y_{it}+y_{iu})} \{1 -[1 +(w-1) (y_{it}+y_{iu})]^{N_I}\} \right].
\end{align}
\end{subequations}
To derive equality \eqref{eq_pi_multinomial}, we used an expression of the mean of the multinomial distribution.
To derive equality \eqref{eq_pi_binomial}, we used a property of the binomial distribution.
To demonstrate the final equality, using the substitutions $y_i \equiv y_{it} +y_{iu}$ and $y_n \equiv y_{nt} +y_{nu}$, we refer to the derivation for the asymmetric NTG in Appendix A of \cite{Lim:2024ty}.
The derivation assumes that $y_i=y_{it}+y_{iu}\ne 0$ and $w\ne 1$,
i.e., $P_i|(y_{it}+y_{iu}\ne 0 \land w\ne 1)$.
For $y_{it}+y_{iu} =0$, we define
$P_i|(y_{it}=y_{iu}=0) \equiv (r+1)y_{nt} -1 =\lim_{y_{it}+y_{iu}\rightarrow 0} P_i|(y_{it}+y_{iu}\ne 0 \land w\ne 1)
$,
applying L'H\^{o}pital's rule.
For $w=1$, we define
$P_i |(w=1) \equiv (r+1) (y_{it}+y_{nt}) -1 =\lim_{w\rightarrow 1} P_i|(y_{it}+y_{iu}\ne 0 \land w\ne 1)
$,
applying L'H\^{o}pital's
 rule.

\subsection{Derivation of Eq.\,\eqref{eq_payoff_trustworthy}
\label{deri_payoff_trustworthy}}

For a trustee in a group of $N$ players with $N_I$ investors and $N_T$ trustees, the probability of having $k_{it}$, $k_{iu}$, $k_{nt}$, and $k_{nu}$ co-players of $it$, $iu$, $nt$, and $nu$ types among the $N_I$ investors is
\begin{equation}
\Pr(k_{it},k_{iu},k_{nt},k_{nu}; N_I) =\frac{N_I!}{k_{it}!k_{iu}!k_{nt}!k_{nu}!} y_{it}^{k_{it}}y_{iu}^{k_{iu}}y_{nt}^{k_{nt}}y_{nu}^{k_{nu}},
\end{equation}
where $k_{it}+k_{iu}+k_{nt}+k_{nu}=N_I$.
The probability of having $l_{it}$, $l_{iu}$, $l_{nt}$, and $l_{nu}$ co-players of $it$, $iu$, $nt$, and $nu$ types among the remaining $N_T-1$ trustees is
\begin{equation}
\Pr(l_{it},l_{iu},l_{nt},l_{nu}; N_T-1) =\frac{(N_T-1)!}{l_{it}!l_{iu}!l_{nt}!l_{nu}!} y_{it}^{l_{it}}y_{iu}^{l_{iu}}y_{nt}^{l_{nt}}y_{nu}^{l_{nu}},
\end{equation}
where $l_{it}+l_{iu}+l_{nt}+l_{nu}=N_T-1$.
For a trustworthy trustee (i.e., $it$ or $nt$ type), the total number of investing investors is $k_{it}+k_{iu}$. 
Thus, the expected payoff $P_t$ for a player acting as a trustworthy trustee is given by
\begin{subequations}
\begin{align}
P_{t} &= \sum_{k_{it}+k_{iu}+k_{nt}+k_{nu}=N_I} \Pr(k_{it},k_{iu},k_{nt},k_{nu}; N_I) \sum_{l_{it}+l_{iu}+l_{nt}+l_{nu}=N_T-1} \Pr(l_{it},l_{iu},l_{nt},l_{nu}; N_T-1) \Pi_t(k_{it}+k_{iu}) \notag \\
&= \sum_{k_{it}+k_{iu}+k_{nt}+k_{nu}=N_I} \frac{N_I!}{k_{it}!k_{iu}!k_{nt}!k_{nu}!} y_{it}^{k_{it}}y_{iu}^{k_{iu}}y_{nt}^{k_{nt}}y_{nu}^{k_{nu}} \sum_{l_{it}+l_{iu}+l_{nt}+l_{nu}=N_T-1} \frac{(N_T-1)!}{l_{it}!l_{iu}!l_{nt}!l_{nu}!} y_{it}^{l_{it}}y_{iu}^{l_{iu}}y_{nt}^{l_{nt}}y_{nu}^{l_{nu}} \frac{r}{N_T} \frac{1-w^{k_{it}+k_{iu}}}{1-w} \notag \\
&= \sum_{k_{it}+k_{iu}+k_{nt}+k_{nu}=N_I} \frac{N_I!}{k_{it}!k_{iu}!k_{nt}!k_{nu}!} y_{it}^{k_{it}}y_{iu}^{k_{iu}}y_{nt}^{k_{nt}}y_{nu}^{k_{nu}} \frac{r}{N_T} \frac{1-w^{k_{it}+k_{iu}}}{1-w} (y_{it}+y_{iu}+y_{nt}+y_{nu})^{N_T-1} \label{eq_pt_multinomial} \\
&= \frac{r}{N_T}\frac{1}{1-w}\sum_{k_{it}+k_{iu}+k_{nt}+k_{nu}=N_I} \frac{N_I!}{k_{it}!k_{iu}!k_{nt}!k_{nu}!} y_{it}^{k_{it}}y_{iu}^{k_{iu}}y_{nt}^{k_{nt}}y_{nu}^{k_{nu}} \left(1-w^{k_{it}+k_{iu}}\right) \notag \\
&= \frac{r}{N_T}\frac{1}{1-w} \left[1 -\sum_{k_{it}+k_{iu}+k_{nt}+k_{nu}=N_I} \frac{N_I!}{k_{it}!k_{iu}!k_{nt}!k_{nu}!} (wy_{it})^{k_{it}}(wy_{iu})^{k_{iu}} y_{nt}^{k_{nt}}y_{nu}^{k_{nu}}\right] \notag \\
&= \frac{r}{N_T}\frac{1}{1-w} \left[1 -(wy_{it}+wy_{iu}+y_{nt}+y_{nu})^{N_I}\right] \notag \\
&= r\frac{1}{N_T(1-w)} \left\{1 -\left[1 +(w-1) (y_{it}+y_{iu})\right]^{N_I}\right\}, 
\end{align}
\end{subequations} 
where $y_{it}+y_{iu}+y_{nt}+y_{nu}=1$ is used. 
To derive equality \eqref{eq_pt_multinomial}, we applied an expression of the multinomial distribution.

\subsection{Derivation of Eq.\,\eqref{eq_payoff_it_investor}\label{deri_eq_payoff_it_investor}}

We express Eq.\,\eqref{eq_payoffs_roles} as
\begin{equation}
\Pi_i(m_i,g_t) =\Pi_{i,1}(m_i,g_t) +\Pi_{i,2}(g_t),
\label{eq_payoff_investing_split}
\end{equation}
where 
$  \Pi_{i,1}(m_i,g_t) \equiv\frac{g_t}{N_T}\frac{r\left(1-w^{m_i}\right)}{m_i(1-w)}$ and
$\Pi_{i,2}(g_t) \equiv \frac{g_t}{N_T} -1$.
To calculate the expection of $\Pi_i$, denoted by $P_{i|it}$, we separately calculate the expected payoffs of $\Pi_{i,1}$ and $\Pi_{i,2}$, denoted by $P_{i|it,1}$ and $P_{i|it,2}$, respectively, and then sum $P_{i|it,1}$ and $P_{i|it,2}$. 

For an investor of $it$ type in a group of $N$ players (with $N_I$ investors and $N_T$ trustees), the probability of having $k_{it}$, $k_{iu}$, $k_{nt}$, and $k_{nu}$ co-players of $it$, $iu$, $nt$, and $nu$ types among the remaining $N_I - 1$ investors obeys the multivariate hypergeometric distribution given by
\begin{equation}
\Pr_{it}(k_{it},k_{iu},k_{nt},k_{nu}; N_I-1) 
=\frac{\binom{N_{it}-1}{k_{it}}\binom{N_{iu}}{k_{iu}}\binom{N_{nt}}{k_{nt}}\binom{N_{nu}}{k_{nu}}}{ \binom{N-1}{N_I-1}},
\end{equation}
where $N_{it}, N_{iu}, N_{nt}$, and $N_{nu}$ denote the number of players of each type; note that $N_{it}+N_{iu}+N_{nt}+N_{nu} =N$ and $k_{it}+k_{iu}+k_{nt}+k_{nu}=N_I-1$. For infinite well-mixed populations, the multinomial distribution is used instead (see Appendix \ref{deri_payoff_investing}).
For an investor of $it$ type, the number of investing investors is $k_{it}+1+k_{iu}$, and the number of trusting trustees is $N_{it}-(k_{it}+1) +N_{nt}-k_{nt}$.
Unlike in well-mixed populations, in structured populations, the numbers of trustees of each type are not independent of $k_{it}$, $k_{iu}$, $k_{nt}$, and $k_{nu}$, but given by $N_{it}-(k_{it}+1), N_{iu}-k_{iu}, N_{nt}-k_{nt}$, and $N_{nu}-k_{nu}$, respectively.
Thus, the expected payoff component $P_{i|it,1} =P_{i|it,1}(N_{it},N_{iu},N_{nt},N_{nu})$ for a player of $it$ type acting as an investing investor is given by
\begin{subequations}   
\begin{align}  
P_{i|it,1} &= \sum_{k_{it}+k_{iu}+k_{nt}+k_{nu}=N_I -1} \Pr_{it}(k_{it},k_{iu},k_{nt},k_{nu}; N_I-1) \Pi_{i,1}(k_{it}+1+k_{iu}, N_{it}-(k_{it}+1) +N_{nt}-k_{nt}) \notag \\
&= \binom{N-1}{N_I-1}^{-1} \sum_{k_{it}+k_{iu}+k_{nt}+k_{nu}=N_I -1} \binom{N_{it}-1}{k_{it}}\binom{N_{iu}}{k_{iu}} \binom{N_{nt}}{k_{nt}}\binom{N_{nu}}{k_{nu}} \notag \\
&\quad \times \frac{\left[N_{it}-(k_{it}+1)+N_{nt}-k_{nt}\right]}{N_T} \frac{r(1-w^{k_{it}+k_{iu}+1})}{(k_{it}+k_{iu}+1)(1-w)} \notag \\
&= \binom{N-1}{N_I-1}^{-1} \frac{r}{N_T(1-w)} \sum_{k_i} \sum_{k_{it}+k_{iu}=k_i} \binom{N_{it}-1}{k_{it}}\binom{N_{iu}}{k_{iu}} \frac{1-w^{k_{it}+k_{iu}+1}}{k_{it}+k_{iu}+1} \notag \\
&\quad \times \sum_{k_{nt}+k_{nu}=N_I -1 -k_i} \binom{N_{nt}}{k_{nt}}\binom{N_{nu}}{k_{nu}} (N_{it}+N_{nt}-k_{it}-1-k_{nt}) \notag \\
&= \binom{N-1}{N_I-1}^{-1} \frac{r}{(N-N_I)(1-w)} \sum_{k_i} \sum_{k_{it}+k_{iu}=k_i} \binom{N_{it}-1}{k_{it}}\binom{N_{iu}}{k_{iu}} \frac{1-w^{k_i+1}}{k_i+1} \notag \\
&\quad \times \left[(N_{it}+N_{nt}-k_{it}-1) \sum_{k_{nt}+k_{nu}=N_I -1 -k_i} \binom{N_{nt}}{k_{nt}}\binom{N_{nu}}{k_{nu}} - \sum_{k_{nt}+k_{nu}=N_I -1 -k_i} \binom{N_{nt}}{k_{nt}}\binom{N_{nu}}{k_{nu}} k_{nt}\right] \notag \\
&= \binom{N-1}{N_I-1}^{-1} \frac{r}{(N-N_I)(1-w)} \sum_{k_i} \frac{1-w^{k_i+1}}{k_i+1} \sum_{k_{it}+k_{iu}=k_i} \binom{N_{it}-1}{k_{it}}\binom{N_{iu}}{k_{iu}} \notag \\
&\quad \times \left[(N_{it}+N_{nt} -k_{it}-1) \binom{N_{nt}+N_{nu}}{N_I -1 -k_i} - \binom{N_{nt}+N_{nu}}{N_I -1 -k_i} (N_I -1 -k_i)\frac{N_{nt}}{N_{nt}+N_{nu}}\right] \label{piit1_Vandermonde_hypergeometric} \\
&= \binom{N-1}{N_I-1}^{-1} \frac{r}{(N-N_I)(1-w)} \sum_{k_i} \binom{N_{nt}+N_{nu}}{N_I -1 -k_i} \frac{1-w^{k_i+1}}{k_i+1} \notag \\
&\quad \times \left[\left\{N_{it}+N_{nt} -1 -(N_I -1 -k_i)\frac{N_{nt}}{N_{nt}+N_{nu}}\right\} \sum_{k_{it}+k_{iu}=k_i} \binom{N_{it}-1}{k_{it}}\binom{N_{iu}}{k_{iu}} - \sum_{k_{it}+k_{iu}=k_i} \binom{N_{it}-1}{k_{it}}\binom{N_{iu}}{k_{iu}}k_{it}\right] \notag \\
&= \binom{N-1}{N_I-1}^{-1} \frac{r}{(N-N_I)(1-w)} \sum_{k_i} \binom{N_{nt}+N_{nu}}{N_I -1 -k_i} \frac{1-w^{k_i+1}}{k_i+1} \notag \\
&\quad \times \left[\left\{N_{it}+N_{nt} -1 -(N_I -1 -k_i)\frac{N_{nt}}{N_{nt}+N_{nu}}\right\} \binom{N_{it}+N_{iu}-1}{k_i} - \binom{N_{it}+N_{iu}-1}{k_i}k_i \frac{N_{it}-1}{N_{it}+N_{iu}-1}\right] \label{piit1_hypergeometric} \\
&= \frac{r}{(1-w)(N-N_I)\binom{N-1}{N_I-1}} \sum_{k_i=0}^{N_I-1} \binom{N_{it}+N_{iu}-1}{k_i} \binom{N_{nt}+N_{nu}}{N_I -1 -k_i} \frac{1-w^{k_i+1}}{k_i+1} \notag \\
&\quad \times \left[N_{it}+N_{nt}-1 -(N_I -1)\frac{N_{nt}}{N_{nt}+N_{nu}} + k_i\left(\frac{N_{nt}}{N_{nt}+N_{nu}} -\frac{N_{it}-1}{N_{it}+N_{iu}-1}\right)\right].
\end{align} 
\end{subequations} 
To derive equality \eqref{piit1_Vandermonde_hypergeometric}, we used expressions of Vandermonde's identity and the mean of the hypergeometric distribution. To derive equality \eqref{piit1_hypergeometric}, we used an expression of the mean of the hypergeometric distribution.  
Note that we assumed $N_{nt}+N_{nu} \ne 0$ when applying an expression of  the mean of the hypergeometric distribution,
$\sum_{k_{nt}+k_{nu}=N_I -1 -k_i}\binom{N_{nt}}{k_{nt}}\binom{N_{nu}}{k_{nu}} k_{nt} =
\binom{N_{nt}+N_{nu}}{N_I -1 -k_i}(N_I -1 -k_i)\frac{N_{nt}}{N_{nt}+N_{nu}}$, to derive equality \eqref{piit1_Vandermonde_hypergeometric}.
If $N_{nt}+N_{nu} = 0$, then the sum equals 0.
We express this sum for both cases where $N_{nt}+N_{nu} \ne 0$ and where $N_{nt}+N_{nu} = 0$ as
\begin{equation}
\sum_{k_{nt}+k_{nu}=N_I -1 -k_i}\binom{N_{nt}}{k_{nt}}\binom{N_{nu}}{k_{nu}} k_{nt} =
\binom{N_{nt}+N_{nu}}{N_I -1 -k_i}(N_I -1 -k_i)\Big\langle\frac{N_{nt}}{N_{nt}+N_{nu}}\Big\rangle_0,
\end{equation}
where
\begin{equation}
\Big\langle\frac{a}{b}\Big\rangle_0 \equiv 
\begin{cases}
\frac{a}{b}, & b\ne 0,\\
0, & b=0.
\end{cases}
\end{equation}
Similarly, we assumed  $N_{it}+N_{iu}-1 \ne 0$ when applying an expression of  the mean of the hypergeometric distribution to derive equality \eqref{piit1_hypergeometric}. If $N_{it}+N_{iu}-1 =0$, then the sum equals 0. We express this sum for both cases where $N_{it}+N_{iu}-1 \ne 0$ and where $N_{it}+N_{iu}-1 =0$ as
\begin{equation}
\sum_{k_{it}+k_{iu}=k_i}
\binom{N_{it}-1}{k_{it}}\binom{N_{iu}}{k_{iu}}k_{it} 
=\binom{N_{it}+N_{iu}-1}{k_i}k_i \Big\langle\frac{N_{it}-1}{N_{it}+N_{iu}-1}\Big\rangle_0.
\end{equation}
 
We have
\begin{subequations} 
\begin{align}
P_{i|it,2} &= \sum_{k_{it}+k_{iu}+k_{nt}+k_{nu}=N_I -1} \Pr_{it}(k_{it},k_{iu},k_{nt},k_{nu}; N_I-1) \Pi_{i,2}(N_{it}-(k_{it}+1) +N_{nt}-k_{nt}) \notag \\
&= \binom{N-1}{N_I-1}^{-1}\sum_{k_{it}+k_{iu}+k_{nt}+k_{nu}=N_I -1} \binom{N_{it}-1}{k_{it}}\binom{N_{iu}}{k_{iu}}\binom{N_{nt}}{k_{nt}}\binom{N_{nu}}{k_{nu}} \left[-1 +\frac{N_{it}-(k_{it}+1)+N_{nt}-k_{nt}}{N-N_I}\right] \notag \\
&= -1+\frac{N_{it}+N_{nt}-1}{N-N_I} -\frac{1}{N-N_I} \binom{N-1}{N_I-1}^{-1} \sum_{k_{it}+k_{iu}+k_{nt}+k_{nu}=N_I -1} \binom{N_{it}-1}{k_{it}}\binom{N_{iu}}{k_{iu}}\binom{N_{nt}}{k_{nt}}\binom{N_{nu}}{k_{nu}} (k_{it}+k_{nt})  \notag\\
&= -1+\frac{N_{it}+N_{nt}-1}{N-N_I} -\frac{1}{N-N_I} \left[(N_I-1)\frac{N_{it}-1}{N-1}+(N_I-1)\frac{N_{nt}}{N-1}\right]  \label{eq_piit2_multivariate_hypergeometric}\\
&= -\frac{N-N_{it}+N_{nt}}{N-1} = -\frac{N_{iu}+N_{nu}}{N-1},
\end{align}
\end{subequations} 
where $N_{it} + N_{iu} + N_{nt} + N_{nu} = N$ is used. To derive equality \eqref{eq_piit2_multivariate_hypergeometric}, we used an expression of the average of the multivariate hypergeometric distribution.
Therefore, the expected payoff $P_{i|it}$ for an $it$ player acting as an (investing) investor is given by
\begin{align}
P_{i|it} &= \sum_{k_{it}+k_{iu}+k_{nt}+k_{nu}=N_I -1} \Pr_{it}(k_{it},k_{iu},k_{nt},k_{nu}; N_I-1) \Pi_i \notag\\
&= \sum_{k_{it}+k_{iu}+k_{nt}+k_{nu}=N_I -1} \Pr_{it}(k_{it},k_{iu},k_{nt},k_{nu}; N_I-1) (\Pi_{i,1}+\Pi_{i,2}) \notag\\
&= P_{i|it,1}+P_{i|it,2} \notag\\
&= -\frac{N_{iu}+N_{nu}}{N-1} + \frac{r}{(1-w)(N-N_I)\binom{N-1}{N_I-1}} \sum_{k_i=0}^{N_I-1} \binom{N_{it}+N_{iu}-1}{k_i} \binom{N_{nt}+N_{nu}}{N_I -1 -k_i} \notag\\
&\quad \times \frac{1-w^{k_i+1}}{k_i+1} \left[N_t-1 -(N_I -1)\Big\langle\frac{N_{nt}}{N_{nt}+N_{nu}}\Big\rangle_0 +k_i\left(\Big\langle\frac{N_{nt}}{N_{nt}+N_{nu}}\Big\rangle_0 -\Big\langle\frac{N_{it}-1}{N_{it}+N_{iu}-1}\Big\rangle_0\right)\right].
\end{align}

\subsection{Derivation of Eq.\,\eqref{eq_payoff_it_trustee}\label{deri_eq_payoff_it_trustee}}

For a trustee of $it$ type in a group of $N$ players (with $N_I$ investors and $N_T$ trustees), the probability of having $k_{it}$, $k_{iu}$, $k_{nt}$, and $k_{nu}$ co-players of $it$, $iu$, $nt$, and $nu$ types among the $N_I$ investors is
\begin{equation}
\Pr_{it}(k_{it},k_{iu},k_{nt},k_{nu}; N_I) 
=\frac{\binom{N_{it}-1}{k_{it}}\binom{N_{iu}}{k_{iu}}\binom{N_{nt}}{k_{nt}}\binom{N_{nu}}{k_{nu}}}{\binom{N-1}{N_I}},
\end{equation}
where $k_{it}+k_{iu}+k_{nt}+k_{nu}=N_I$.
For a trustee of $it$ type, the number of investing investors in the group is $k_{it}+k_{iu}$.
Thus, the expected payoff $P_{t|it}$ for an $it$ player acting as a (trustworthy) trustee is given by
\begin{subequations} 
\begin{align}
P_{t|it} &= \sum_{k_{it}+k_{iu}+k_{nt}+k_{nu}=N_I} \Pr_{it}(k_{it},k_{iu},k_{nt},k_{nu}; N_I) \Pi_t(k_{it}+k_{iu}) \notag \\
&= \binom{N_{it}-1+N_{iu}+N_{nt}+N_{nu}}{N_I}^{-1} \sum_{k_{it}+k_{iu}+k_{nt}+k_{nu}=N_I} \binom{N_{it}-1}{k_{it}}\binom{N_{iu}}{k_{iu}} \binom{N_{nt}}{k_{nt}}\binom{N_{nu}}{k_{nu}} \Pi_t(k_{it}+k_{iu}) \notag \\
&= \binom{N-1}{N_I}^{-1} \sum_{k_i} \sum_{k_{it}+k_{iu}=k_i} \sum_{k_{nt}+k_{nu}=N_I -k_i} \binom{N_{it}-1}{k_{it}}\binom{N_{iu}}{k_{iu}} \binom{N_{nt}}{k_{nt}}\binom{N_{nu}}{k_{nu}} \Pi_t(k_i) \notag \\
&= \binom{N-1}{N_I}^{-1} \sum_{k_i} \Pi_t(k_i) \sum_{k_{it}+k_{iu}=k_i} \binom{N_{it}-1}{k_{it}}\binom{N_{iu}}{k_{iu}} \sum_{k_{nt}+k_{nu}=N_I -k_i} \binom{N_{nt}}{k_{nt}}\binom{N_{nu}}{k_{nu}} \notag \\
&= \binom{N-1}{N_I}^{-1} \sum_{k_i} \Pi_t(k_i) \binom{N_{it}+N_{iu}-1}{k_i} \binom{N_{nt}+N_{nu}}{N_I -k_i}
\label{eq_ptit_Vandermonde_1}\\
&= \binom{N-1}{N_I}^{-1} \sum_{k_i} \binom{N_{it}+N_{iu}-1}{k_i}\binom{N_{nt}+N_{nu}}{N_I -k_i} \frac{r}{N-N_I} \frac{1-w^{k_i}}{1-w} \notag \\
&= \binom{N-1}{N_I}^{-1} \frac{r}{(N-N_I)(1-w)} \left[\sum_{k_i}\binom{N_{it}+N_{iu}-1}{k_i}\binom{N_{nt}+N_{nu}}{N_I -k_i} - \sum_{k_i}\binom{N_{it}+N_{iu}-1}{k_i}\binom{N_{nt}+N_{nu}}{N_I -k_i}w^{k_i}\right] \notag \\
&= \binom{N-1}{N_I}^{-1} \frac{r}{(N-N_I)(1-w)} \left[\binom{N-1}{N_I} -\sum_{k_i}\binom{N_{it}+N_{iu}-1}{k_i}\binom{N_{nt}+N_{nu}}{N_I -k_i} w^{k_i}\right] \label{eq_ptit_Vandermonde_2}\\
&= \frac{r}{(1-w)(N-N_I)} \left[1 -\frac{1}{\binom{N-1}{N_I}}\sum_{k_i=0}^{N_I} \binom{N_{it}+N_{iu}-1}{k_i}\binom{N_{nt}+N_{nu}}{N_I -k_i}w^{k_i}\right].
\end{align}
\end{subequations} 
To derive equalities \eqref{eq_ptit_Vandermonde_1} and \eqref{eq_ptit_Vandermonde_2}, we used an expression of Vandermonde's identity.

\subsection{Derivation of Eq.\,\eqref{eq_payoff_iu_investor}, \eqref{eq_payoff_iu_trustee}, \eqref{eq_payoff_nt_trustee}, and \eqref{eq_payoff_nu_trustee}\label{deri_eq_payoff_iu_nt_nu}}

The derivation of $P_{i|iu}$ follows that of $P_{i|it}$ in Appendix \ref{deri_eq_payoff_it_investor}, but replaces $N_{it}$ with $N_{it}+1$ and $N_{iu}$ with $N_{iu}-1$.
The differences between $P_{i|iu}$ and $P_{i|it}$ lie in the probability:
$\Pr_{iu}(\cdot) =\binom{N_{it}}{k_{it}}\binom{N_{iu}-1}{k_{iu}}\binom{N_{nt}}{k_{nt}}\binom{N_{nu}}{k_{nu}}/ \binom{N-1}{N_I-1}$ versus 
$\Pr_{it}(\cdot) =\binom{N_{it}-1}{k_{it}}\binom{N_{iu}}{k_{iu}}\binom{N_{nt}}{k_{nt}}\binom{N_{nu}}{k_{nu}}/ \binom{N-1}{N_I-1}$
and the number of trustworthy trustees:
$N_{it}-k_{it} +N_{nt}-k_{nt}$ versus $N_{it}-(k_{it}+1) +N_{nt}-k_{nt}$.
We can derive $P_{i|iu}$ from $P_{i|it}$ by making these replacements.
The expected payoff $P_{i|iu}$ for an $iu$ player acting as an (investing) investor is then given by
\begin{align}
P_{i|iu} = &-\frac{N_{iu}+N_{nu}-1}{N-1} + \frac{r}{(1-w)(N-N_I)\binom{N-1}{N_I-1}} \sum_{k_i=0}^{N_I-1} \binom{N_{it}+N_{iu}-1}{k_i} \binom{N_{nt}+N_{nu}}{N_I -1 -k_i} \nonumber \\
&\times \frac{1-w^{k_i+1}}{k_i+1} \left[N_t -(N_I -1) \Big\langle\frac{N_{nt}}{N_{nt}+N_{nu}}\Big\rangle_0 +k_i\left(\Big\langle\frac{N_{nt}}{N_{nt}+N_{nu}}\Big\rangle_0 -\Big\langle\frac{N_{it}}{N_{it}+N_{iu}-1}\Big\rangle_0\right)\right].
\label{eq_line_break}
\end{align}

The derivation of $P_{u | iu}$ follows that of $P_{t|it}$ in Appendix \ref{deri_eq_payoff_it_trustee}, differing by the scale factor $\frac{1}{r}$. Another difference lies in the probability:
$\Pr_{it}(\cdot) 
=\binom{N_{it}-1}{k_{it}}\binom{N_{iu}}{k_{iu}}\binom{N_{nt}}{k_{nt}}\binom{N_{nu}}{k_{nu}}/ \binom{N-1}{N_I}$
versus
$\Pr_{iu}(\cdot) 
=\binom{N_{it}}{k_{it}}\binom{N_{iu}-1}{k_{iu}}\binom{N_{nt}}{k_{nt}}\binom{N_{nu}}{k_{nu}}/ \binom{N-1}{N_I}$.
However, this distinction does not cause any difference in the derivation of $P_{u | iu}$ and $P_{t|it}$, because the crucial factor is the number of $it$ and $iu$ co-players in the group, which is $N_{it}+N_{iu}-1$ in both cases.
Therefore, the expected payoff $P_{u|iu} (N_{it},N_{iu},N_{nt},N_{nu})$ for a player of $iu$ type acting as an (untrustworthy) trustee is:
\begin{align}
P_{u|iu} (N_{it},N_{iu},N_{nt},N_{nu})
=& \sum_{k_{it}+k_{iu}+k_{nt}+k_{nu}=N_I}\Pr_{iu}(k_{it},k_{iu},k_{nt},k_{nu}; N_I) \Pi_u(k_{it}+k_{iu}) \notag\\
=& \frac{1}{r}P_{t|it} (N_{it},N_{iu},N_{nt},N_{nu}).
\end{align}

The derivation of $P_{t|nt}$ is similar to that of $P_{t|it}$ in Appendix \ref{deri_eq_payoff_it_trustee}, but replaces $N_{it}$ with $N_{it}+1$ and $N_{nt}$ with $N_{nt}-1$, because the difference lies in the probability:
$\Pr_{nt}(\cdot) =\binom{N_{it}}{k_{it}}\binom{N_{iu}}{k_{iu}}\binom{N_{nt}-1}{k_{nt}}\binom{N_{nu}}{k_{nu}}/ \binom{N-1}{N_I}$
versus
$\Pr_{it}(\cdot) =\binom{N_{it}-1}{k_{it}}\binom{N_{iu}}{k_{iu}}\binom{N_{nt}}{k_{nt}}\binom{N_{nu}}{k_{nu}}/ \binom{N-1}{N_I}$.
Thus, the expected payoff $P_{t|nt}$ for an $nt$ player acting as a trustworthy trustee is:
\begin{align}
P_{t|nt} &= \sum_{k_{it}+k_{iu}+k_{nt}+k_{nu}=N_I} \Pr_{nt}(k_{it},k_{iu},k_{nt},k_{nu}; N_I) \Pi_t(k_{it}+k_{iu}) \notag \\
&=\frac{r}{(1-w)(N-N_I)} \left[1 -\frac{1}{\binom{N-1}{N_I}}\sum_{k_i=0}^{N_I} \binom{N_{it}+N_{iu}}{k_i}\binom{N_{nt}+N_{nu}-1}{N_I -k_i} w^{k_i}\right].
\end{align}

Using a derivation similar to that of $P_{u|iu} (\cdot)
= \frac{1}{r}P_{t|it} (\cdot)$ above,
we obtain
\begin{equation}
P_{u|nu} (N_{it},N_{iu},N_{nt},N_{nu})
=\frac{1}{r} P_{t|nt}(N_{it},N_{iu},N_{nt},N_{nu}).
\end{equation}

\subsection{Equilibria at Vertices\label{proof_vertices}}

To analyse the dynamics given by Eq.\,\eqref{eq_population_dynamics}, we find all equilibria by solving $\dot{y}_{it} = \dot{y}_{iu} =\dot{y}_{nt}  =\dot{y}_{nu} = 0$. We assess local stability of each equilibrium by the sign of the eigenvalues of the Jacobian matrix, $J$, at the equilibrium. An equilibrium is stable if it has no positive eigenvalues; it is unstable otherwise. The Jacobian is a $3 \times 3$ matrix due to the constraint $y_{it}+y_{iu}+y_{nt}+y_{nu} =1$. 
For equilibria at vertices or edges of the simplex, stability analysis can be simplified. At a vertex, eigenvectors align with the edges. On an edge, one eigenvector aligns with the edge, whereas the other two eigenvectors lie in the adjacent triangular faces \cite{Priklopil:2017aa}.

\subsubsection{IT, IU and NT vertices, $y_{it}=1$, $y_{iu}=1$, $y_{nt}=1$}

The equilibrium at $y_{it}=1$ is unstable. We show this by considering the dynamics along the edge $y_{it}+y_{iu}=1$, given by
$\dot{y}_{it} =y_{it}(1 -y_{it}) \tanh \left(\frac{1}{2} \beta  (P_{it}-P_{iu})\right)$.        
The eigenvalue of the $1 \times 1$ Jacobian matrix $\pdv{\dot{y}_{it}}{y_{it}}$ at the equilibrium is $\pdv{\dot{y}_{it}}{y_{it}}\big|_{y_{it}=1} =\frac{(1-r) \left(w^{N_I}-1\right)}{N (w-1)}>0$.
Therefore, the equilibrium is unstable.

Similarly,  at $y_{iu}=1$, the eigenvalue for the direction of $y_{iu}+y_{nu}=1$ is $\pdv{\dot{y}_{iu}}{y_{iu}}\big|_{y_{iu}=1} =\frac{N_I}{N}>0$.
At $y_{nt}=1$, the eigenvalue for the direction of $y_{it}+y_{nt}=1$ is $\pdv{\dot{y}_{nt}}{y_{nt}}\big|_{y_{nt}=1} =\frac{N_I r}{N} >0$.

\subsubsection{NU  vertex, $y_{nu}=1$}
        
The equilibrium at vertex NU is stable, which we show as follows.
The Jacobian at $y_{nu}=1$ is
\begin{equation} 
\begin{split}
J_{NU} &= \left(
\begin{array}{ccc}
\pdv{\dot{y}_{it}}{y_{it}} &
\pdv{\dot{y}_{it}}{y_{iu}} &
\pdv{\dot{y}_{it}}{y_{nt}} 
\\
\pdv{\dot{y}_{iu}}{y_{it}} &
\pdv{\dot{y}_{iu}}{y_{iu}} &
\pdv{\dot{y}_{iu}}{y_{nt}} 
\\
\pdv{\dot{y}_{nt}}{y_{it}} &
\pdv{\dot{y}_{nt}}{y_{iu}} &
\pdv{\dot{y}_{nt}}{y_{nt}} 
\end{array}
\right)
\\
& = \left(
\begin{array}{ccc}
J_{11 \mid NU}  & 0 & 0 \\
0 & J_{22 \mid NU}  & 0 \\
0 & 0 & 0 \\
\end{array}
\right),
\end{split}
\end{equation}
where $J_{11 | NU} =J_{22 | NU} = -\tanh \left(\frac{\beta  N_I}{2 N}\right) <0$.
The eigenvalues are $\lambda_1 =\lambda_2 = J_{11 | NU}<0$, and $\lambda_3=0$.
Therefore, the equilibrium is stable.

\subsection{Equilibria on Edges \label{proof_edges}}

\subsubsection{The interior of IT-IU edge, $y_{it}+y_{iu}=1$}

No equilibrium exists in the interior of the IT-IU edge because
$P_{it}-P_{iu} <0$ when $y_{it}+y_{iu}=1$ and $0 <y_{it} <1$.
An equilibrium would require $P_{it}-P_{iu} =0$.

\subsubsection{IT-NT edge, $y_{it}+y_{nt}=1$} 

Similarly, there is no equilibrium in the interior of the IT-NT edge because
$P_{it}-P_{nt} >0$ when $y_{it}+y_{nt}=1$ and $0 <y_{it} <1$.

\subsubsection{IT-NU edge, $y_{it}+y_{nu}=1$\label{proof_equil_IT-NU}}  

A unique unstable equilibrium exists in the interior of the IT-NU edge.

\textit{(Proof of existence and uniqueness of the equilibrium):}
Given $y_{it}+y_{nu}=1$ and $0<y_{it}<1$,
we have
$P_{it}-P_{nu} 
=\frac{1}{N}\left[\frac{(2 r-1) \left(\left[(w-1) y_{it}+1\right]^{N_I}-1\right)}{w-1}+N_I (y_{it}-1)\right]
$.
There exists a unique $y_{it}^* \in (0,1)$ such that $(P_{it}-P_{nu})|_{y_{it}=y_{it}^*} =0$.
This equality holds true because $P_{it}-P_{nu}$ is continuous and strictly increases with $y_{it}$, 
$(P_{it}-P_{nu})|_{y_{it}=0} =-\frac{N_I}{N} <0$
and $(P_{it}-P_{nu})|_{y_{it}=1} =\frac{(2 r-1) \left(w^{N_I}-1\right)}{N (w-1)} >0$.
We assume $r>\frac{1}{2}$ to ensure that the NTG is a social dilemma.
Quantity $P_{it}-P_{nu}$ strictly increases with $y_{it}$ 
because $\pdv{y_{it}}(P_{it}-P_{nu})  =\frac{N_I (2 r-1) \left[(w-1) y_{it}+1\right]^{N_I-1}+N_I}{N} >0$,
which follows from $N_I >0$, $N>0$, $2 r-1 >0$ , $w>0$, $0<y_{it} <1$, and $(w-1) y_{it}+1 \ge 0$.

\textit{(Proof of instability):}
The equilibrium at $y_{it}=y_{it}^*$ is unstable. We show this by considering the dynamics along the edge $y_{it}+y_{iu}=1$.
The eigenvalue of the $1 \times 1$ Jacobian matrix $\pdv{\dot{y}_{it}}{y_{it}}$ at the equilibrium is 
$\pdv{\dot{y}_{it}}{y_{it}} \big|_{y_{it}=y^*_{it}} 
=y_{it}(1-y_{it})\pdv{y_{it}}(P_{it}-P_{nu}) >0
$.

\subsubsection{IU-NT edge, $y_{iu}+y_{nt}=1$\label{proof_IU-NT}}

Any equilibrium $\mathbf{R}$ in the interior of the IU-NT edge is unstable.
It suffices to show instability in a subspace of the state space. We shall prove this in the subspace spanned by the $iu$, $nt$ and $nu$ strategies.
In this subspace, we obtain
\begin{subequations} 
\begin{align}
\dot{y}_{iu} &=y_{iu} \left[(1-y_{iu}-y_{nu}) \tanh \left(\frac{1}{2} \beta 
   (P_{iu}-P_{nt})\right) 
  +y_{nu} \tanh \left(\frac{1}{2} \beta  (P_{iu}-P_{nu})\right)\right], \\
\dot{y}_{nu} &=y_{nu} \left[y_{iu} \tanh \left(\frac{1}{2} \beta 
   (P_{nu}-P_{iu})\right)
  +(1-y_{iu}-y_{nu}) \tanh \left(\frac{1}{2} \beta  (P_{nu}-P_{nt})\right)\right].
\end{align}
\end{subequations} 
The Jacobian at  $\mathbf{R}$ is given by
\begin{equation}
J_{\mathbf{R}} =
\left(
\begin{array}{cc}
J_{11|\mathbf{R}}  & J_{12|\mathbf{R}}  \\
0 & J_{22|\mathbf{R}} 
\end{array}
\right),
\end{equation}
where
$J_{22 |\mathbf{R}} =\pdv{\dot{y}_{nu}}{y_{nu}} \big|_\mathbf{R}  
=y_{iu} \tanh \left(\frac{1}{2} \beta 
   (P_{nu}-P_{iu})\right)+(1-y_{iu}) \tanh \left(\frac{1}{2} \beta  (P_{nu}-P_{nt})\right) >0
$.       
Note that $J_{22 |\mathbf{R}} >0$ holds true at $\mathbf{R}$ because $0<y_{iu}<1$, $P_{nu} -P_{iu} >0$ and $P_{nu} -P_{nt} >0$. Inequality $P_{nu} -P_{nt} >0$ holds true because combination of Eqs. \eqref{eq_payoff_untrustworthy} and \eqref{eq_expected_payoffs} yields $P_{nu} -P_{nt} =(1-P_I)(P_u -P_t) >0$. Furthermore, $P_{nu} -P_{iu} >0$ follows from $P_{nu} -P_{nt} >0$ and $P_{nt} =P_{iu}$, with the latter equality following from the definition of equilibrium $\mathbf{R}$ on the IU-NT edge. The eigenvalues are $\lambda_{1} =J_{11 |\mathbf{R}}$ and $\lambda_{2} =J_{22 |\mathbf{R}}$. The equilibrium $\mathbf{R}$ is unstable since $\lambda_{2} >0$.

\subsubsection{IU-NU edge, $y_{iu}+y_{nu}=1$}

No equilibrium exists on the IU-NU edge because $P_{iu}-P_{nu} <0$ when $y_{iu}+y_{nu}=1$ and $0<y_{iu}<1$.

\subsubsection{NT-NU edge,  $y_{nt} +y_{nu} =1$\label{stability_NT-NU}}  

The Jacobian at a point on this edge is
\begin{align}
J_{(NT-NU)} 
  &= \left(
\begin{array}{ccc}
J_{11 | (NT-NU)} & 0 & 0 \\
 0 & J_{22 | (NT-NU)} & 0 \\
J_{31 | (NT-NU)} & J_{32 | (NT-NU)} & 0 \\
\end{array}
\right),
\end{align}
where $J_{11 | (NT-NU)}=J_{22 | (NT-NU)}= \tanh \left(\frac{\beta  N_I [(r+1)y_{nt}-1]}{2 N}\right)$.
The eigenvalues are $\lambda_{1} =0$ and $\lambda_{2} =\lambda_{3} =J_{11 | (NT-NU)}$.
For $y_{nt} <\frac{1}{r+1}$, the equilibrium is stable because $\lambda_{1}=0$ and $\lambda_{2} =\lambda_{3} <0$.
For $y_{nt} >\frac{1}{r+1}$, the equilibrium is unstable because $\lambda_{2} =\lambda_{3} >0$.
In other words, the segment $\frac{r}{r+1}< y_{nu}< 1$ is stable, while $0 <y_{nu} <\frac{r}{r+1}$ is unstable, given $y_{nt}+y_{nu}=1$.

\subsection{No Equilibrium on the Faces \label{proof_faces}}

In this section, we show that no interior equilibrium exists on the four faces of the simplex $\triangle^3$.

\subsubsection{The interior of the IT-IU-NT face, $y_{it}>0$, $y_{iu}>0$, $y_{nt}>0$, and $y_{nu}=0$\label{proof_no_equilibrium_IT-IU-NT}}

No equilibrium exists in the interior of the IT-IU-NT face ($y_{it}>0$, $y_{iu}>0$, $y_{nt}>0$, and $y_{nu}=0$).
We prove this by contradiction.
If an equilibrium existed, we would have $\dot{y}_{it}=\dot{y}_{iu}=\dot{y}_{nt}=\dot{y}_{nu}=0$ at that point, 
yielding 
 $\dv{t}  \left(\frac{y_{iu}}{y_{it}}\right)  =\frac{y_{iu}}{y_{it}}\left(\frac{\dot{y}_{iu}}{y_{iu}} -\frac{\dot{y}_{it}}{y_{it}}\right) =0$,
which contradicts $\dv{t}  \left(\frac{y_{nu}}{y_{nt}}\right) >0$.
Thus, no equilibrium exists.

\textit{(Proof of $\dv{t}  \left(\frac{y_{iu}}{y_{it}}\right) >0$):}
Because
\begin{equation}
\dv{t}  \left(\frac{y_{iu}}{y_{it}}\right) =\frac{y_{it} \dot{y}_{iu}-y_{iu} \dot{y}_{it}}{y_{it}^2} =\frac{y_{iu}}{y_{it}}\left(\frac{\dot{y}_{iu}}{y_{iu}} -\frac{\dot{y}_{it}}{y_{it}}\right),
\end{equation}
$y_{it}>0$ and $y_{iu}>0$,
it follows that
\begin{equation}
\text{Sign}\left(\dv{t}  \left(\frac{y_{iu}}{y_{it}}\right)\right) =\text{Sign}\left(\frac{\dot{y}_{iu}}{y_{iu}} -\frac{\dot{y}_{it}}{y_{it}}\right).
\end{equation}
To prove that $\dv{t} \left(\frac{y_{iu}}{y_{it}}\right) >0$,
therefore,
it suffices to show $\frac{\dot{y}_{iu}}{y_{iu}} -\frac{\dot{y}_{it}}{y_{it}}>0$.
We have
\begin{equation}
\frac{\dot{y}_{iu}}{y_{iu}} -\frac{\dot{y}_{it}}{y_{it}}=  \frac{2 y_{nt} e^{\beta  P_{nt}}  \left(e^{\beta  P_{iu}}-e^{\beta  P_{it}}\right)}{\left(e^{\beta  P_{it}}+e^{\beta  P_{nt}}\right) \left(e^{\beta  P_{iu}}+e^{\beta 
   P_{nt}}\right)}      +(y_{it}+y_{iu}) \tanh \left(\frac{1}{2} \beta  (P_{iu}-P_{it})\right) >0
\end{equation}
since $P_{iu}-P_{it}>0$, given $y_{it}+y_{iu}>0$.
Thus, we have proven $\dv{t}  \left(\frac{y_{iu}}{y_{it}}\right) >0$.

\subsubsection{The interior of the IT-IU-NU, IT-NT-NU, and IU-NT-NU faces}

No interior equilibrium exists in the interior of
the remaining three faces. For IT-IU-NU, a proof
similar to that of IT-IU-NT in Appendix
\ref{proof_no_equilibrium_IT-IU-NT} applies
because $P_{iu}-P_{it}>0$ holds true in the interior of both faces.
For IT-NT-NU, an interior equilibrium would lead
to $\dot{y}_{nt}=\dot{y}_{nu}=0$, contradicting
$\dv{t} \left(\frac{y_{nu}}{y_{nt}}\right) >0$.
We obtain $\dv{t} \left(\frac{y_{nu}}{y_{nt}}\right) >0$ because
\begin{equation}
\frac{\dot{y}_{nu}}{y_{nu}} -\frac{\dot{y}_{nt}}{y_{nt}} =
(e^{\beta P_{nu}}-e^{\beta P_{nt}}) \left[ \frac{2 y_{it} e^{\beta P_{it}}}{(e^{\beta P_{it}}+e^{\beta P_{nt}})(e^{\beta P_{it}}+e^{\beta P_{nu}})}
+\frac{y_{nt}+y_{nu}}{e^{\beta P_{nt}}+e^{\beta P_{nu}}} \right] > 0,
\end{equation}
which holds true under $P_{nu} >P_{nt}$ and $y_{it}>0$.
The proof for IU-NT-NU resembles that for IT-NT-NU
because $P_{nu} >P_{nt}$ holds true in the interior of both faces.

\subsection{No Equilibrium Inside the 3-Simplex \label{proof_simplex}}

No equilibrium exists in the interior of the IT-IU-NT-NU simplex (i.e., $y_{it}>0$, $y_{iu}>0$, $y_{nt}>0$, and $y_{nu}>0$). The proof of this is similar to the previous proofs. An interior equilibrium would lead to $\dot{y}_{it}=\dot{y}_{iu}=0$, contradicting $\dv{t}  \left(\frac{y_{iu}}{y_{it}}\right) >0$. We obtain $\dv{t}  \left(\frac{y_{iu}}{y_{it}}\right) >0$ because
\begin{align}
\frac{\dot{y}_{iu}}{y_{iu}} -\frac{\dot{y}_{it}}{y_{it}} =& 2 \left(e^{\beta  P_{iu}}-e^{\beta  P_{it}}\right) \left[\frac{y_{nt} e^{\beta  P_{nt}}}{\left(e^{\beta  P_{it}} + e^{\beta  P_{nt}}\right) \left(e^{\beta  P_{iu}}+e^{\beta  P_{nt}}\right)}  +\frac{y_{nu} e^{\beta  P_{nu}}}{\left(e^{\beta  P_{it}}+e^{\beta  P_{nu}}\right) \left(e^{\beta  P_{iu}}+e^{\beta  P_{nu}}\right)}\right] \notag \\
& +(y_{it}+y_{iu}) \tanh \left(\frac{1}{2} \beta  (P_{iu}-P_{it})\right) >0.
\label{eq:no-eq-inside-simplex-a}
\end{align}
Equation~\eqref{eq:no-eq-inside-simplex-a} holds true because $P_{iu}-P_{it}>0$ given $y_{it}+y_{iu}>0$.

\subsection{Derivation of the threshold $r^*$ for the square lattice \label{deri_threshold_r_homo}}

For the $it$ and $nu$ players on the border between the $it$ and $nu$ clusters in the infinite square lattice (see Fig.\,\ref{fig_analysis_IT_NU_homogeneous}(a)), $P_{it}-P_{nu}>0$ holds true if and only if $r>r^*$. Here, $r=r^*$ is the unique solution of $P_{it}-P_{nu}=0$, given $w$.

A player on a grid belongs to five groups: one centred on itself and one for each of its four immediate neighbours. Each group comprises five players. A player participates in a 5-player TG associated with each group.
For the $it$ and $nu$ players on the border, we obtain
\begin{equation}
P_{it} = P_{it}(N_{it}=1,N_{nu}=4) +3P_{it}(N_{it}=4,N_{nu}=1) + P_{it}(N_{it}=5)
\end{equation}
and
\begin{equation}
P_{nu} =P_{nu}(N_{nu}=5) +3P_{nu}(N_{it}=1,N_{nu}=4) +P_{nu}(N_{it}=4,N_{nu}=1).
\end{equation}

For $p_I=3/5$ (i.e., $N_I=3$),
we obtain
\begin{equation}
P_{it} = \frac{1}{20} \{2 r [w (7 w+16)+16]-21\}
\end{equation}
and
\begin{equation}
P_{nu} = \frac{1}{20} [4 w (w+1)+13].
\end{equation}
By solving
\begin{equation}
P_{it}-P_{nu} =\frac{1}{10} \{r [w (7 w+16)+16]-2 w (w+1)-17\} =0
\end{equation}
for $r$, we obtain
\begin{equation}
r = r^* =\frac{2 w (w+1)+17}{w (7 w+16)+16}.
\end{equation}
We conclude that $P_{it}-P_{nu}>0$ if and only if $r>r^*$ because $P_{it}-P_{nu}$ is a linear function of $r$ and $(P_{it}-P_{nu})|_{r=0}=\frac{1}{10} \left[-2 w (w+1)-17\right]<0$.

Similarly, for $p_I =1/5$,
we have
$P_{it}-P_{nu} =\frac{1}{10} (13 r-7)$ and $r^*=\frac{7}{13}$.
For $p_I =2/5$, we have
$P_{it}-P_{nu} =\frac{1}{5} [r (5 w+8)-w-6]$ and $r^*=\frac{w+6}{5 w+8}$.
For $p_I =4/5$, we have
$P_{it}-P_{nu} =\frac{1}{5} \left\{2 r \left[w (w+2)^2+4\right] -w \left(w^2+w+1\right)-11\right\}$ and  $r^* =\frac{w^3+w^2+w+11}{2 \left(w^3+4 w^2+4 w+4\right)}$.

\subsection{Derivation of Eqs.\,\eqref{eq_threshold_hetero} and \eqref{eq_threshold_hetero_infinity} \label{deri_threshold_r_hetero}}
 
We have
\begin{align}
P_{it}|\text{hub} &= P_{it}|\text{(centred at the $it$ hub)} + (N^{(2)}-2)P_{it}|\text{(centred at the $it$ leaf)} + P_{it}|\text{(centred at the $nu$ hub)} \notag \\
&= \left[p P_{i|it}(N_{it}=N^{(2)}-1,N_{nu}=1) + (1-p)P_{t|it}(N_{it}=N^{(2)}-1,N_{nu}=1)\right]  \notag \\
& \quad + (N^{(2)}-2) [p P_{i|it}(N_{it}=2) + p P_{t|it}(N_{it}=2)]  \notag \\
&\quad 
+ [pP_{i|it}(N_{it}=1,N_{nu}=N^{(1)}-1) + (1-p) P_{t|it}(N_{it}=1,N_{nu}=N^{(1)}-1)] \notag \\
&= \left[p\left(\frac{r w^{N_I}+w}{w-N^{(2)} w}+\frac{r (w^{N_I}-1)}{N_I (w-1)}\right) - (1-p)\frac{r \left(\frac{N_I (w-1) w^{N_I-1}}{N^{(2)}-1}-w^{N_I}+1\right)}{(w-1) (N^{(2)} -N_I)}\right] 
+ (N^{(2)}-2) [pr +(1-p)r] \notag \\
&\quad +[p \cdot (-1) +(1-p) \cdot 0] \notag \\
&\approx \frac{w [N^{(2)} (p-pw -2r)+2r] -2r w^{N^{(2)} p} [N^{(2)} (p-1) w-N^{(2)} p+w]}{(N^{(2)}-1) N^{(2)} (w-1) w} + (N^{(2)}-2) r-p,
\end{align}
where we omitted any of $N_{it}=0,N_{iu}=0,N_{nt}=0$, or $N_{nu}=0$ from the argument of the probability,
e.g., $P_{i | it}(N_{it}=N^{(2)}-1,N_{nu}=1) \equiv$
$P_{i | it}(N_{it}=N^{(2)}-1,N_{nu}=1,N_{iu}=N_{nt}=0)$. We also used 
$N_I = \lceil N^{(2)}p \rceil \approx N^{(2)}p$ for
the number of investors in the group centred at the $it$ hub.

We have
\begin{align}
P_{nu}|\text{hub} &= P_{nu}|\text{(centred at the $nu$ hub)} + (N^{(1)}-2)P_{nu}|\text{(centred at the $nu$ leaf)} + P_{nu}|\text{(centred at the $it$ hub)} \notag \\
&= [p P_{n|nu}(N_{it}=1,N_{nu}=N^{(1)}-1) + (1-p) P_{u|nu}(N_{it}=1,N_{nu}=N^{(1)}-1)] \nonumber \\
&\quad + (N^{(1)}-2)[p P_{n|nu}(N_{nu}=2) + (1-p)P_{u|nu}(N_{nu}=2)] \nonumber \\
&\quad + [p P_{n|nu}(N_{it}=N^{(2)}-1) + (1-p)P_{u|nu}(N_{it}=N^{(2)}-1)] \notag \\
&= \left[ p \cdot 0 + (1-p) \frac{N_I^\prime}{(N^{(1)}-1) (N^{(1)}-N_I^\prime)} \right] + (N^{(1)}-2)[p \cdot 0 + (1-p) \cdot 0] + \left[ p \cdot 0 + (1-p)\frac{w^{N_I}-1}{(w-1) (N^{(2)}-N_I)}\right] \nonumber \\
&\approx (1-p) \frac{N^{(1)}p}{(N^{(1)}-1) (N^{(1)}-N^{(1)}p)} + (1-p)\frac{w^{N^{(2)}p}-1}{(w-1) (N^{(2)}-N^{(2)}p)} \notag \\ 
&= \frac{p}{N^{(1)}-1} + \frac{w^{N^{(2)} p}-1}{N^{(2)} (w-1)},
\end{align}
where we used $N_I^\prime = \lceil N^{(1)}p \rceil \approx N^{(1)}p$ and $N_I = \lceil N^{(2)}p \rceil \approx N^{(2)}p$,
for the numbers of investors in the groups centred at the $nu$ and $it$ hubs, respectively.

By solving $ P_{it}| \text{hub}  -P_{nu}| \text{hub} =0$ for $r$, we obtain
\begin{equation}
r^* = \frac{\text{num}}{\text{denom}},
\end{equation}
where
\begin{align}
\text{num} \equiv& w \left[ N^{(1)} \left(N^{(2)}\right)^2 p w - N^{(1)} \left(N^{(2)}\right)^2 p + (N^{(1)}-1) (N^{(2)}-1) w^{N^{(2)} p} - N^{(1)} N^{(2)} \right. \notag\\
& \left. + N^{(1)} - N^{(2)} p w + N^{(2)} p + N^{(2)} - 1 \right]
\end{align}
and
\begin{equation}
\text{denom}  \equiv (N^{(1)}-1) \left\{(N^{(2)}-1) w \left[ (N^{(2)}-2) N^{(2)} (w-1) - 2 \right] - 2 w^{N^{(2)} p} \left[ N^{(2)} (p-1) w - N^{(2)} p + w \right] \right\}.
\end{equation}
For $0<w <1$, we obtain
\begin{subequations} 
\begin{align}
\lim_{N^{(1)} \to \infty, \, N^{(2)} \to \infty} r^* 
=&    \lim_{N^{(1)} \to \infty, \, N^{(2)} \to \infty} \frac{\text{num}}{\text{denom}}  \notag\\
=& \lim_{N^{(1)} \to \infty, \, N^{(2)} \to \infty} \frac{w \left[N^{(1)} \left(N^{(2)}\right)^2 p (w - 1)  - N^{(1)} (N^{(2)} -1) - N^{(2)} (p w +p +1)  - 1 + w^{N^{(2)} p} O\left(N^{(1)}N^{(2)}\right) \right]}{ (N^{(1)}-1) \left\{(N^{(2)}-1) w \left[ (N^{(2)}-2) N^{(2)} (w-1) - 2 \right]  - 2 w^{N^{(2)} p}O\left(N^{(1)}N^{(2)}\right)\right\}}  \notag\\
=& \lim_{N^{(1)} \to \infty, \, N^{(2)} \to \infty} \frac{w \left[N^{(1)} \left(N^{(2)}\right)^2 p (w - 1)   - N^{(1)} (N^{(2)} -1) - N^{(2)} (p w +p +1)  - 1 \right]}{ (N^{(1)}-1) (N^{(2)}-1) w \left[ (N^{(2)}-2) N^{(2)} (w-1) - 2 \right] }
\label{eq_exp_decay}\\
=& \frac{O\left(N^{(1)}\left(N^{(2)}\right)^2\right)}{O\left(N^{(1)}\left(N^{(2)}\right)^3\right)}\notag\\
=& 0, 
\end{align}
\end{subequations} 
where $O(\cdot)$ denotes the order of the function.
To derive equality \eqref{eq_exp_decay}, we used the observation that the exponential decay of $w^{N^{(2)} p} \to 0$ dominates over the polynomial growth of $N^{(1)} N^{(2)} \to \infty$ as $ N^{(1)} \to \infty$ and $ N^{(2)} \to \infty$.

For $w =1$, both the numerator and denominator of $r^*$ are 0. Therefore, using L'H\^{o}pital's
rule, we obtain
\begin{align}
r^*
=& \lim_{w\rightarrow 1} \frac{\text{num}(w)}{\text{denom}(w)} \notag\\
=& \lim_{w\rightarrow 1} \frac{\text{num}^\prime(w)}{\text{denom}^\prime(w)} \notag\\
=& \frac{p [N^{(1)} (2 N^{(2)}-1)-N^{(2)}]}{(N^{(1)}-1) (N^{(2)}-2) (N^{(2)}+2 p-1)} \notag\\
=& \frac{O\left(N^{(1)}N^{(2)}\right)}{O\left(N^{(1)}\left(N^{(2)}\right)^2\right)}. 
\end{align}
Therefore, $\lim_{N^{(1)} \to \infty, \, N^{(2)} \to \infty} r^* =0$.
 
For $w>1$,
we have
\begin{subequations} 
\begin{align}
\lim_{N^{(1)} \to \infty, \, N^{(2)} \to \infty} r^* 
=&    \lim_{N^{(1)} \to \infty, \, N^{(2)} \to \infty} \frac{\text{num}}{\text{denom}}  \notag\\
=& \lim_{N^{(1)} \to \infty, \, N^{(2)} \to \infty} \frac{w\left[w^{N^{(2)}p}(N^{(1)}-1) (N^{(2)}-1) +O\left(N^{(1)}\left(N^{(2)}\right)^2\right)\right]}{ (N^{(1)}-1) \left\{- 2 w^{N^{(2)} p} \left[ N^{(2)} (p-1) w - N^{(2)} p + w \right]  +O\left(N^{(1)}\left(N^{(2)}\right)^3\right)\right\}}  \notag\\
=& \lim_{N^{(1)} \to \infty, \, N^{(2)} \to \infty} \frac{w\left[w^{N^{(2)}p}(N^{(1)}-1) (N^{(2)}-1)\right]}{ (N^{(1)}-1) \left\{- 2 w^{N^{(2)} p} \left[ N^{(2)} (p-1) w - N^{(2)} p + w \right]\right\}}  \label{eq_exp_growth}\\
=& \lim_{N^{(1)} \to \infty, \, N^{(2)} \to \infty} \frac{w  (N^{(2)}-1)}{- 2\left[ N^{(2)} (p-1) w - N^{(2)} p + w \right]}  \notag \\
=& \lim_{N^{(1)} \to \infty, \, N^{(2)} \to \infty} \frac{w  (1-1/N^{(2)})}{- 2\left[(p-1) w - p + w/N^{(2)} \right]}  \notag \\
=& \frac{w}{- 2\left[(p-1) w - p\right]}  \notag \\
=& \frac{w}{2\left[ p+(1-p) w\right]}.
\end{align}
\end{subequations} 
To derive equality \eqref{eq_exp_growth}, we used the observation that the exponential growth of $w^{N^{(2)} p} \to \infty$ dominates over the polynomial growth of both $N^{(1)}\left(N^{(2)}\right)^2 \to \infty$ and $N^{(1)}\left(N^{(2)}\right)^3 \to \infty$ as $ N^{(1)} \to \infty$ and $ N^{(2)} \to \infty$.

\subsection{Robustness}

\begin{figure*}[t!]  
\begin{center}  
\includegraphics[width=0.94\textwidth]{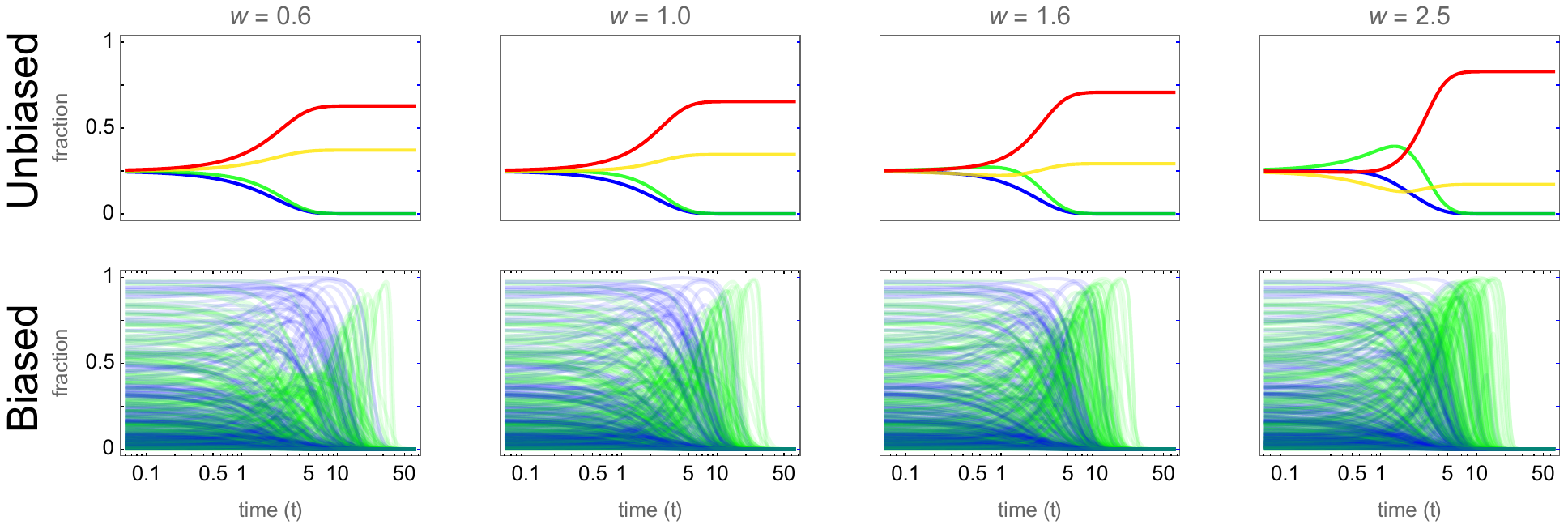}    \\
\includegraphics[width=0.2\textwidth]{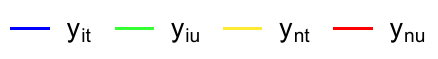}  
\end{center}   
\caption{
Time courses of the SNTG in the infinite well-mixed population with initial conditions in the interior of state space $\triangle^3$.
The fractions of the different strategies, $y_{it}(t)$, $y_{iu}(t)$, $y_{nt}(t)$, and $y_{nu}(t)$ are shown as a function of time for various initial conditions.
(Top row)
Unbiased initial state $y_{it}(0)=y_{iu}(0)=y_{nt}(0)=y_{nu}(0)=1/4$. 
(Bottom) 256 randomly selected initial conditions. We only show $y_{it}(t)$ and $y_{iu}(t)$ for clarity.
We observe $y_{it}(t) +y_{iu}(t) \rightarrow0$ as $t\rightarrow \infty$, regardless of the initial condition;
in other words, $y_{nt}(t) +y_{nu}(t) \rightarrow 1$ as $t\rightarrow \infty$.
Hence, investment does not evolve.
Parameters are the same as those used in Fig.\,\ref{fig_well_mixed_simplex}.}  
\label{fig_well_mixed}   
\end{figure*}

\begin{figure*}[t!]  
\begin{center}  
\includegraphics[width=0.94\textwidth]{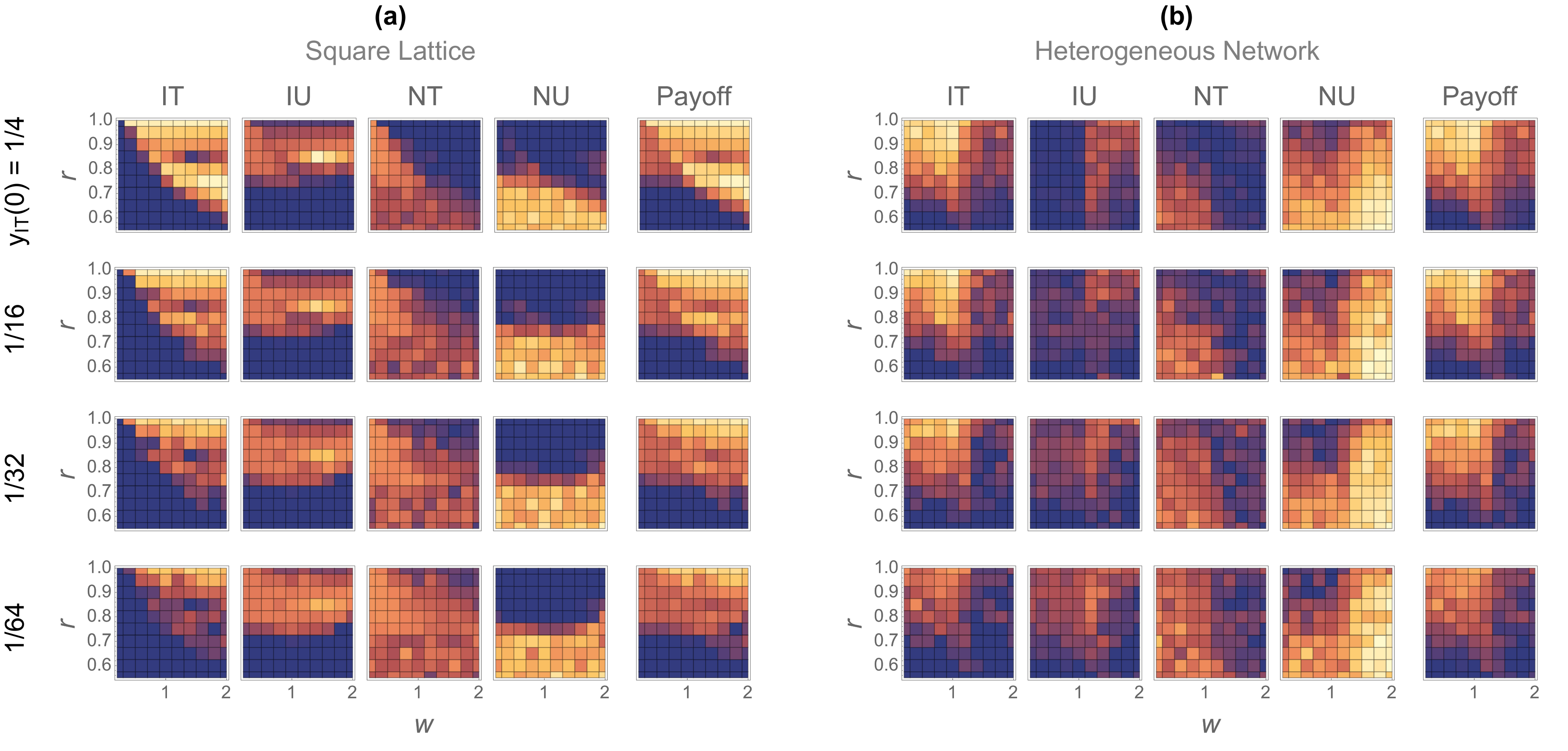}       
\includegraphics[width=0.23\textwidth]{symmetric_NTG_Fig_unit_32}  
\end{center}   
\caption{ 
Robustness of evolutionary outcomes to intial condition $\left(y_{it}(0), y_{iu}(0), y_{nt}(0), y_{nu}(0)\right)$, where $y_{iu}(0) = y_{nt}(0) =y_{nu}(0) =\frac{1}{3} \left(1-y_{it}(0)\right)$.
Changes in the initial condition produce outcomes qualitatively similar to those observed under the baseline condition, $y_{it}(0)=y_{iu}(0)=y_{nt}(0)=y_{nu}(0)=1/4$.
This robustness holds for the tested initial conditions which are biased against the most prosocial type, IT (i.e., beginning with a lower initial proportion of IT compared to the baseline). 
Note that outcomes remain qualitatively similar even when the initial proportion $y_{it}(0)$ of IT is reduced by over an order of magnitude (e.g., from 1/4 to 1/64).
}
\label{fig_initial_condition} 
\end{figure*}

\begin{figure*}[t!]  
\begin{center}  
\includegraphics[width=0.94\textwidth]{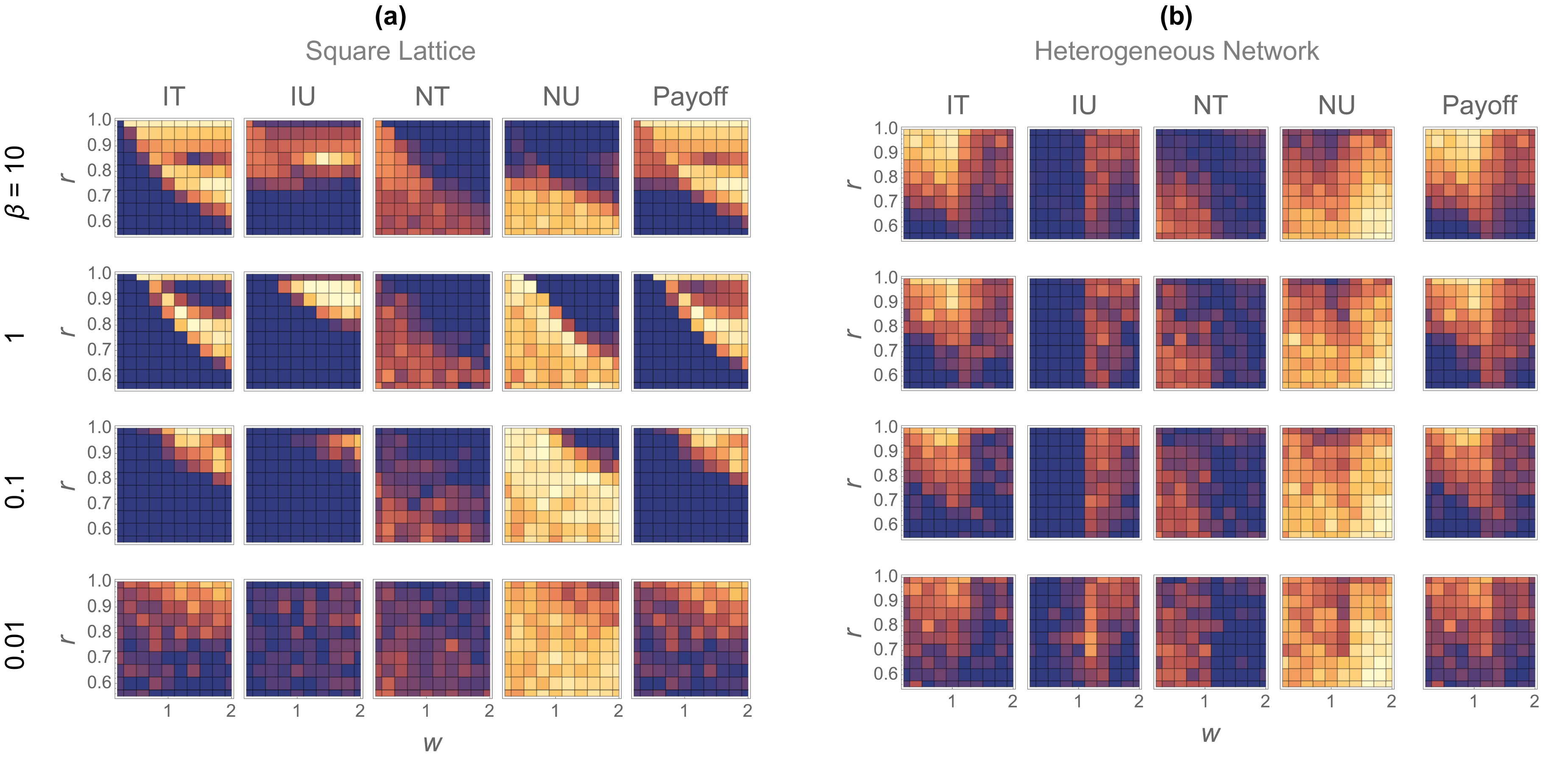}     
\includegraphics[width=0.23\textwidth]{symmetric_NTG_Fig_unit_32}  
\end{center}   
\caption{ 
Robustness of evolutionary outcomes to variations in selection strength, $\beta$. Changes in selection strength yield outcomes that are qualitatively similar to those observed under the baseline condition ($\beta=10$). Note that  outcomes remain qualitatively similar even when the selection strength is reduced by two orders of magnitude (e.g., from 10 to 0.1). 
The range of parameters favourable to the evolution of IT narrows as $\beta$ decreases to 0.1, and then broadens with further decreases in $\beta$, particularly on the square lattice. 
}
\label{fig_selection_strength} 
\end{figure*}

\begin{figure*}[t!]  
\begin{center}  
\includegraphics[width=0.94\textwidth]{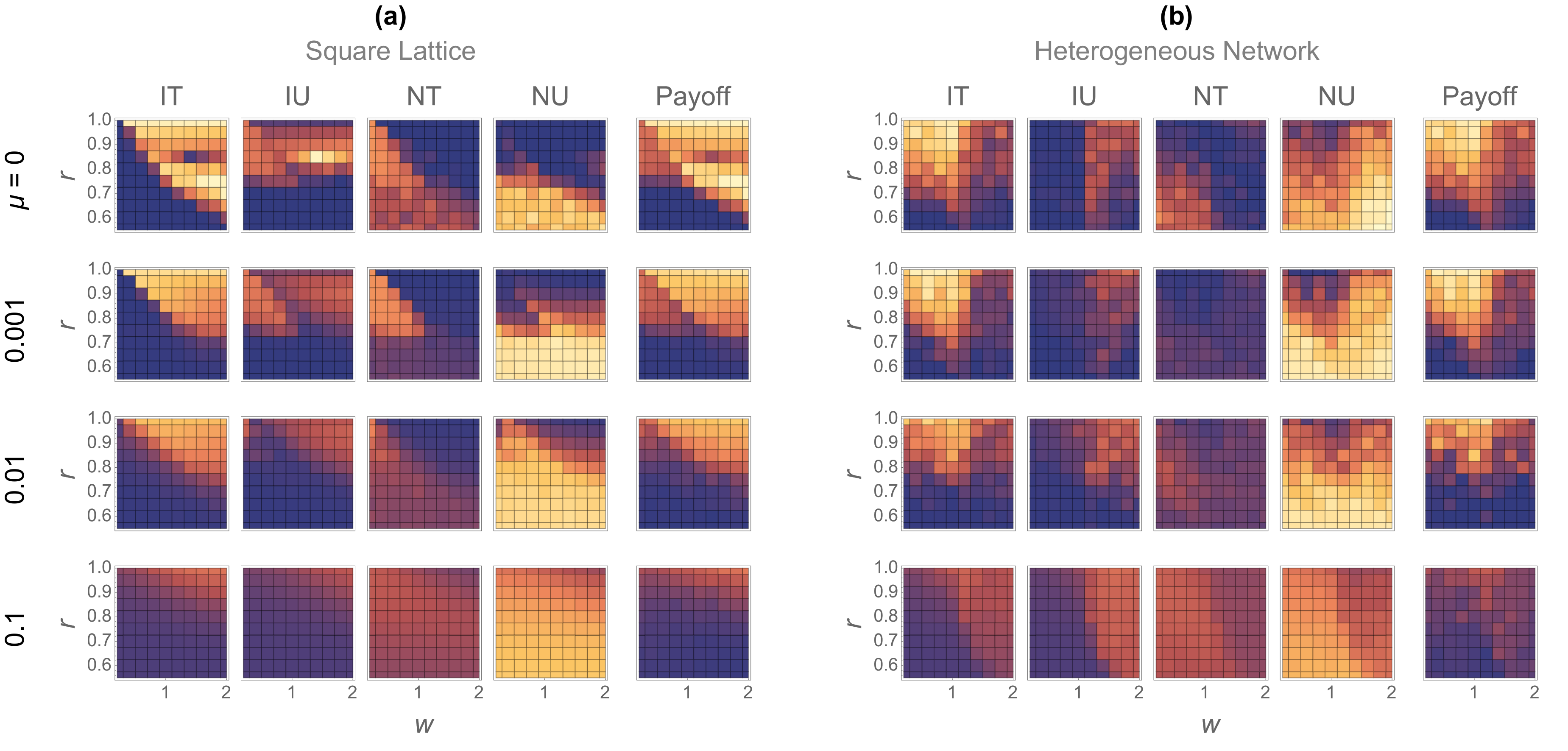}     
\includegraphics[width=0.23\textwidth]{symmetric_NTG_Fig_unit_32}  
\end{center}   
\caption{ 
Robustness of evolutionary outcomes to the mutation rate, $\mu$. Variations in the mutation rate yield outcomes qualitatively similar to those observed under the baseline condition of no mutation ($\mu=0$). Note that outcomes remain qualitatively similar even when the mutation rate is increased tangibly from zero (e.g., up to $\mu=0.01$).
}
\label{fig_mutation_rate} 
\end{figure*}

\begin{figure*}[t!]  
\begin{center}  
\includegraphics[width=0.94\textwidth]{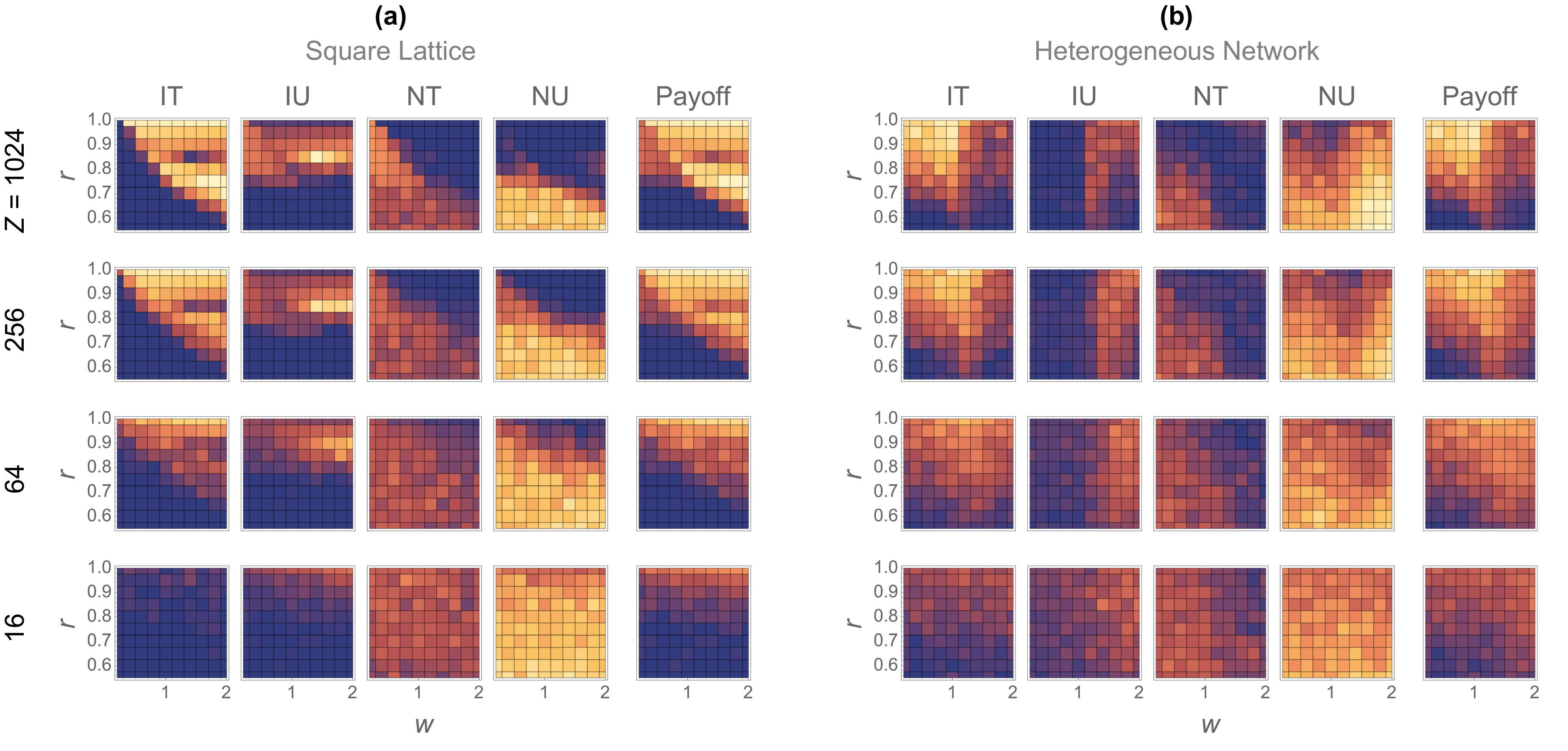}  
\includegraphics[width=0.23\textwidth]{symmetric_NTG_Fig_unit_32}  
\end{center}   
\caption{ 
Robustness of evolutionary outcomes to the population size, $Z$. Variations in population size yield outcomes qualitatively similar to those observed under the baseline condition ($Z=1024$). Note that outcomes remain qualitatively similar even when $Z$ is substantially reduced (e.g., from $Z=1024$ to $Z=64$).
}
\label{fig_population_size} 
\end{figure*}

\begin{figure*}[t!]  
\begin{center}  
\includegraphics[width=0.94\textwidth]{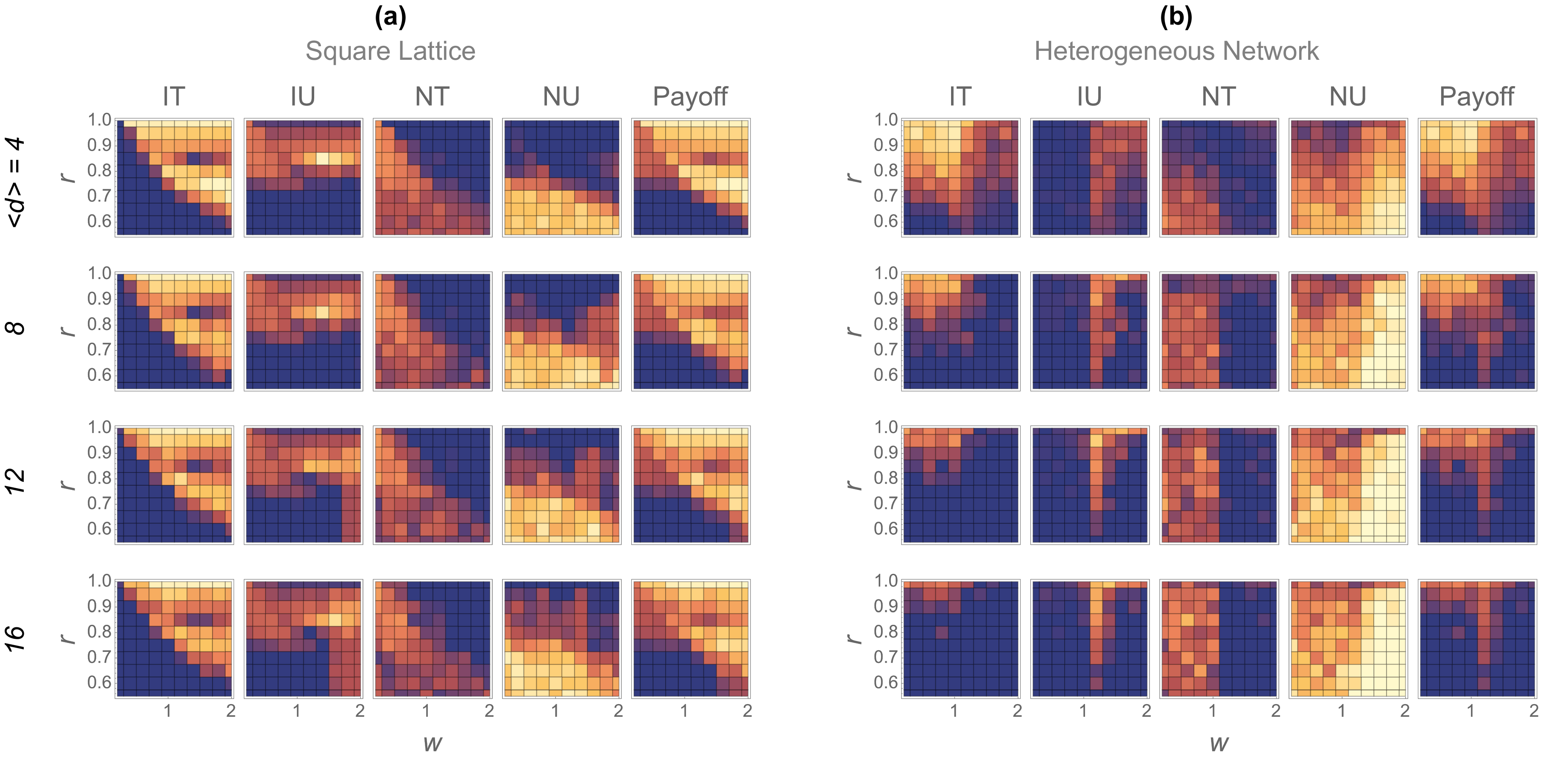}       
\includegraphics[width=0.23\textwidth]{symmetric_NTG_Fig_unit_32}  
\end{center}   
\caption{ 
Robustness of evolutionary outcomes to variations in the mean node degree, $\langle d\rangle$. Changes in the mean node degree yield outcomes qualitatively similar to those observed under the baseline condition ($\langle d\rangle=4$). 
Importantly, the key distinctions in outcomes between the square lattice and heterogeneous networks remain well-preserved across these variations. 
However, as $\langle d\rangle$ increases, the parameter range conducive to the evolution of IT narrows more markedly in heterogeneous networks compared to square lattices.
}
\label{fig_node_degree} 
\end{figure*}

\bibliographystyle{IEEEtran}    


\end{document}